\documentclass[a4paper]{article} 

\usepackage[utf8]{inputenc}
\usepackage[T1]{fontenc}
\usepackage{mathptmx}          
\usepackage{amsmath}
\usepackage{amssymb}
\usepackage{amsfonts}
\usepackage{comment}

\usepackage{graphicx}
\usepackage{float}
\usepackage{booktabs}
\usepackage{tabularx}
\usepackage{multirow}
\usepackage{multicol}
\usepackage{array}
\usepackage{longtable}
\usepackage{caption}
\usepackage{subcaption}
\usepackage{pdflscape}

\usepackage{setspace}
\usepackage{parskip}
\usepackage{enumitem}
\usepackage{fancyhdr}
\usepackage{titlesec}
\usepackage{authblk}
\usepackage{lineno}

\usepackage{xcolor}

\usepackage{xurl} 
\usepackage[
  colorlinks = true,
  linkcolor  = blue,
  citecolor  = blue,
  urlcolor   = blue
]{hyperref}

\usepackage{tikz}
\usetikzlibrary{shapes.geometric, arrows.meta, positioning, backgrounds, fit, calc}
\usepackage{geometry}
\usepackage{lmodern}
\usepackage{fontawesome5}

\definecolor{myred}{RGB}{192,0,0}
\definecolor{myorange}{RGB}{230,115,0}
\definecolor{myyellow}{RGB}{220,180,0}
\definecolor{mygreen}{RGB}{0,150,80}
\definecolor{myblue}{RGB}{0,90,180}
\definecolor{mypurple}{RGB}{120,0,180}
\definecolor{mydarkred}{RGB}{130,0,0}
\definecolor{mylightblue}{RGB}{173,216,230}
\definecolor{mylightgreen}{RGB}{144,238,144}
\definecolor{mylightyellow}{RGB}{255,255,153}
\definecolor{mylightorange}{RGB}{255,200,120}
\definecolor{mylightpurple}{RGB}{220,180,255}
\definecolor{mylightred}{RGB}{255,160,160}
\definecolor{boxgray}{RGB}{245,245,245}

\tikzset{
  blockR/.style={rectangle, rounded corners=6pt,
                 draw=#1!80!black, fill=#1!25,
                 text width=5.2cm, align=center,
                 minimum height=1.1cm, font=\large},
  arrow/.style={-{Stealth[length=7pt]}, line width=1.4pt, color=#1},
  titlebox/.style={rectangle, rounded corners=4pt,
                   draw=#1!80!black, fill=#1!60,
                   text=white, font=\bfseries\Large,
                   minimum height=0.9cm, align=center},
  smallblock/.style={rectangle, rounded corners=4pt,
                     draw=#1!80!black, fill=#1!20,
                     text width=3.2cm, align=center,
                     minimum height=0.9cm, font=\large},
  wideblock/.style={rectangle, rounded corners=6pt,
                    draw=#1!80!black, fill=#1!20,
                    text width=11cm, align=center,
                    minimum height=0.9cm, font=\large},
}

\usepackage{siunitx}
\usepackage{textcomp}
\usepackage{gensymb}
\usepackage{upgreek}
\usepackage{eurosym}

\usepackage[journal=jacsat, articletitle=true, doi=false, super=false, biblabel=brackets, maxauthors=15]{achemso}
\setcitestyle{square,comma,numbers,sort&compress}

\begin{document}
\title{New Avenues of Space Sustainability: Insurance, Liability, and Market Mechanisms for Orbital Debris Remediation}

\author[a]{Junhao Chen}
\author[b,c]{Partha Sarathi Pal}
\author[b*]{Meng Su}
\author[b]{SeyedAbdolreza Sadjadi}
\author[b]{Andreas Ritter}


\affil[a]{Eastside Preparatory School, 10613 NE 38th Pl, Kirkland, WA USA}
\affil[b]{Laboratory for Space Research, FoS, The University of Hong Kong, Hong~Kong~SAR, China}
\affil[c]{Institute of Astronomy Space and Earth Science, Kolkata, India}
\affil[*]{Address correspondence to: mengsu84@hku.hk}

\maketitle

\noindent
\begin{center}
\small
\textit{Received: DD Month YYYY} \quad
\textit{Revised: DD Month YYYY} \quad
\textit{Accepted: DD Month YYYY} \quad
\textit{Published: DD Month YYYY}
\end{center}

\begin{abstract}
Orbital debris is the classic unpriced externality of the NewSpace era. The costs of collision risk, spectrum interference, and lost operational capability are socialized across the orbital commons, while the cost of remediation is borne privately. This article provides a technical and commercial assessment of the debris observation and remediation industry and asks whether, and under what institutional design, space sustainability can become a profitable line of business rather than a regulatory burden. It reviews the state of the art in space situational awareness and space domain awareness (SSA/SDA), e.g., ground-based phased-array radar, optical and space-based sensing, cataloging and conjunction analysis, and the current taxonomy of active debris removal (ADR) and end-of-life (EOL) technologies, from rigid and flexible capture to contactless deorbit. It then maps the commercial landscape that has emerged since 2015, across SSA data-as-a-service providers, remediation and on-orbit servicing firms, and the space insurance sector. The legal architecture, with the 1967 Outer Space Treaty, the 1972 Liability Convention, and national licensing and indemnification authorities, frames these markets. The central argument is that remediation suffers from a public-goods and weakest-link problem that pure commercial demand cannot resolve. Durable sustainability requires a coupled incentive architecture in which regulation (binding deorbit mandates), measurement (commercial SSA with forensic attribution capability), finance (insurance-linked warranties, disposal deposits, and performance bonds), and liability reform (apportionment and attribution) jointly convert remediation from a liability into an asset. The article concludes with concrete technical, commercial, and policy recommendations, including a Hong Kong SAR perspective on orbital insurance and sustainability finance.
\end{abstract}

\noindent\rule{\linewidth}{0.8pt}


\vspace{0.3cm}
\noindent\textbf{Keywords:}
orbital debris; space situational awareness; active debris removal; space insurance; liability; commercial space sustainability; on-orbit servicing.

\vspace{0.5cm}




\section{Introduction}\label{sec1} 
\begin{figure}
\centering
\includegraphics[width=0.45\linewidth]{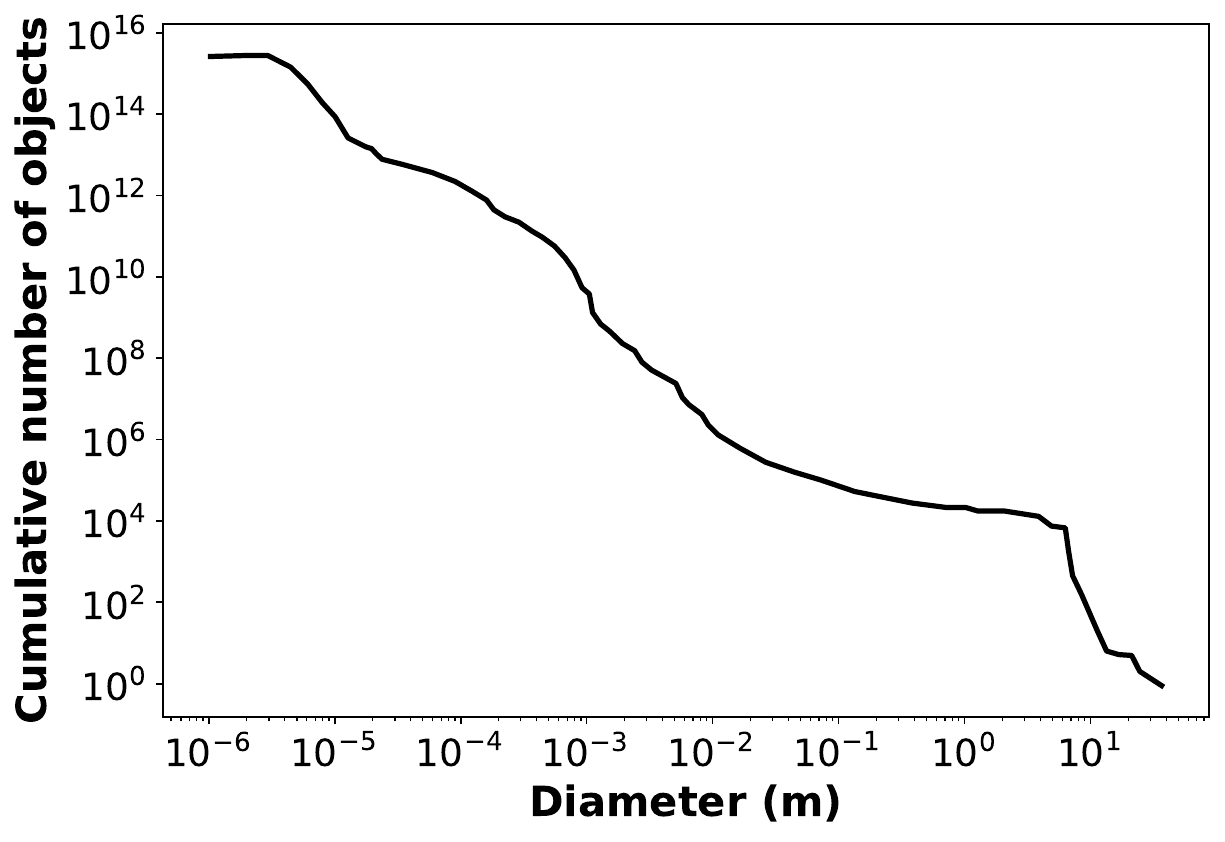}
\caption{Estimated population of space debris in LEO as a function of the size from the
February 2026 reference population. Data credit: \texttt{ESA’s} annual space environment report \url{https://www.sdo.esoc.esa.int/publications/Space_Environment_Report_I10R1_20260908.pdf}.}
\label{deb_size}
\end{figure}

The space age has established itself in science, technology, and, more importantly, society after almost 70 years. It has already changed lifestyles so quickly that a stark contrast can be seen between the present and a decade ago. Such accelerated developments are now gradually facing their consequences, such as space debris. 
With the frequent launch of commercial and non-commercial space vehicles, the Low Earth orbit has become too congested with space debris. This has become an emerging risk or threat to humanity.
Space surveillance networks routinely track and catalog around 47,280 objects, of which only about 15,800 are working satellites \cite{esa2026, ep2026}. The rest consists of abandoned hardware and fragments, and this count covers only objects large enough to be routinely tracked. \texttt{ESA's MASTER-8} population model (reference population February 2026) puts the population above 10 cm at roughly 68,450 objects, the $1-10~cm$ band at about 1.5 million, and the $1~mm-1~cm$ band on the order of 230 million; total on-orbit mass now exceeds 17,400 tonnes. These numbers grow on their own. In calendar year 2024 alone, fragmentation events generated more than 3,000 newly cataloged fragments, and \texttt{ESA's} annual space environment report projects that collision rates will grow exponentially if current behaviour continues \cite{esa2025}. Figure.~\ref{deb_size} shows a cumulative estimate of the listed space debris above the threshold size crossing LEO, based on population statistics up to February 2026.

The increasing amount of space debris in LEO has heightened concerns about space safety for space operators. In its semi-annual filing to the FCC covering June through November 2025, \texttt{SpaceX} reported 148,696 collision-avoidance maneuvers — 44,559 by first-generation Starlink satellites and 104,137 by second-generation spacecraft. This is increased by roughly 50,000 from the comparable reporting period a year earlier \cite{kt2026}. Each maneuver costs propellant, reduces mission margin, and adds uncertainty to the shared catalog. Above a critical density, collisional cascading — the Kessler syndrome will generate debris faster than natural decay removes it, regardless of whether anyone launches again \cite{1978JGR....83.2637K}.

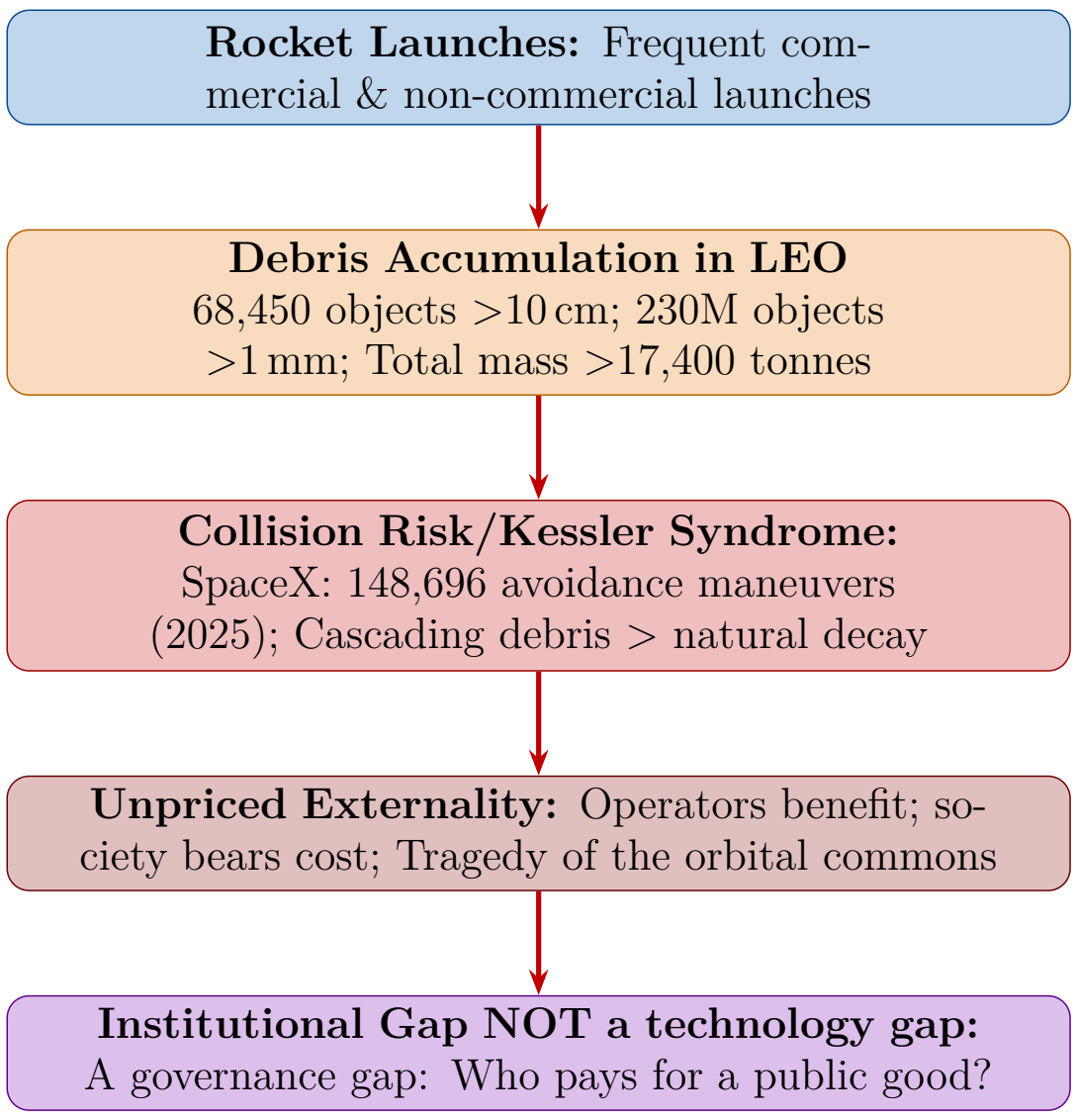
\begin{figure}[htbp]
\centering
\resizebox{0.4\textwidth}{!}{

\begin{tikzpicture}[node distance=1.0cm]

  \node[blockR=myblue, text width=10cm] (A) 
  {{\bfseries Rocket Launches:} Frequent commercial \& non-commercial launches};

  \node[blockR=myorange, text width=10cm, below=of A] (B)
    {{\bfseries Debris Accumulation in LEO}\\
     68{,}450 objects $>$10\,cm;\ 230M objects $>$1\,mm;
     Total mass $>$17{,}400 tonnes};

  \node[blockR=myred, text width=10cm, below=of B] (C)
    {\textbf{Collision Risk/Kessler Syndrome:}\\
     SpaceX: 148{,}696 avoidance maneuvers (2025);
     Cascading debris $>$ natural decay};

  \node[blockR=mydarkred, text width=10cm, below=of C] (D)
    {{\bfseries Unpriced Externality:} Operators benefit; society bears cost;
     Tragedy of the orbital commons};

  \node[blockR=mypurple, text width=10cm, below=of D] (E)
    {\textbf{Institutional Gap NOT a technology gap:} \\ A governance gap:
     Who pays for a public good?};

  \foreach \s/\t in {A/B, B/C, C/D, D/E}{
    \draw[arrow=myred] (\s) -- (\t);
  }


\end{tikzpicture}}
\caption{Space Debris Problem Chain}
\label{fl2}
\end{figure}

Orbital debris is, at its core, an externality. Operators enjoy the benefits of launch and operation, but the collision and interference risk they create is shared by everyone else. The orbital environment is a common resource, as in Hardin's tragedy of the commons, and space debris remediation is purely for the public interest. It reduces the collision risk for all, and these are non-excludable and non-rivalrous, which is why private markets systematically ignore them \cite{1968Sci...162.1243H, 1998Sci...280..682H, weeden2012}. On top of that sits a weakest-link structure. The safety of the orbital environment is set by its least responsible operator. Even perfect compliance by the leading actors cannot secure the commons.
The remediation gap is therefore not primarily a technology gap. The sensing and removal base has matured — commercial-grade inspections of large debris are in orbit right now. The gap is institutional. Who pays for a public good when the beneficiaries cannot be billed, and how does the payment have to be structured for remediation to become a 
profitable line of business? 
The Figure~\ref{fl2} summarizes the space debris issue as a diagram.

\section{Space Debris Observation \& Monitoring} \label{sec2}
Different space debris tracking methods have been developed and tested. These methods are discussed below:

\subsection{Ground-based radar} 
Ground-based radar remains the main instrument for LEO surveillance because it is weather-independent and provides direct range and velocity observation and tracking. There are two types of Radars used for this purpose. Monostatic narrow-beam radars such as the \texttt{MIT Haystack-class instruments} and the \texttt{German FGAN TIRA} achieve high sensitivity and precise observations, but they are constrained by limited dwell time. On the other hand, phased-array radars provide persistent, wide-area surveillance. \texttt{LeoLabs} operates the largest private space-surveillance radar network, with \texttt{AESA (Active Electronically Scanned Array)}, 3-D pulse-Doppler Tracker radars at fixed sites in Texas, New Zealand, Costa Rica, the Azores and Western Australia, plus the transportable Scout-S unit offered for direct sale to allied governments \cite{nse2026a}. \texttt{LeoLabs'} three radar classes (Tracker, Seeker and Scout) trade off sensitivity, field of regard and mobility. National systems, including the U.S. Space Surveillance Network (run by the Space Force's 18th and 19th Space Defense Squadrons), the \texttt{French GRAVES} radar, and \texttt{ESA's SST} network, remain the authoritative public backbone, but commercial radar has become the marginal-data provider for LEO.

\subsection{Optical \& space-based observation} 
Optical sensors dominate the MEO/GEO, and high-altitude LEO, where radar geometry is not useful. \texttt{ExoAnalytic} operates the world's largest private optical telescope network, providing persistent GEO coverage and maneuver detection. \texttt{Slingshot Aerospace} combines an optical sensor network with digital mission engineering and astrodynamics platforms, including orbital collision screening and ``digital twin'' simulations of space operations. Space-based sensing adds a fundamentally different advantage. The \texttt{NorthStar Earth \& Space} deploys optical sensors in orbit to observe LEO, MEO, and GEO objects that ground sensors cannot reliably track, and by April 2026 had four satellites in orbit, with a public-market debut in late 2026 signaling capital-market validation of space-domain awareness (SDA) as an asset class \cite{nse2026b}. Satellite laser ranging stations provide centimeter-level range precision for calibration and tracking high-value objects.

\subsection{Catalogs, conjunction analysis \& the AI applications} 
The public catalog maintained by the \texttt{Combined Space Operations Center (CSpOC)} is the baseline for collision avoidance worldwide. Commercial providers differentiate on catalog completeness, custody (persistent tracking of specific objects), and data quality. \texttt{LeoLabs'} catalog was jointly licensed in December 2025 by the \texttt{U.S. Department of Commerce's Office of Space Commerce} and the \texttt{U.S. Space Force for space safety and security missions}, which is an early institutional endorsement of commercial space situational awareness(SSA) as public infrastructure \cite{leolab2025}. Conjunction assessment pipelines use sensor tracks, propagate ephemerides with realistic covariance, sort against the catalog, and issue conjunction data messages (CDMs). The quality criteria depend on whether uncertainty estimates are optimistic or stale, which can cause misses to become collision events. Using machine learning, these issues can be resolved. Pattern-of-life analysis, maneuver detection from astrometric residuals, anomaly detection in sensor tasking, and AI-driven sensor scheduling are operations typically performed using machine learning. Firms such as \texttt{Karman+}, specialize in multi-source data fusion, and \texttt{Privateer} provides open-data analytics. The 2026 IAF Space Traffic Management working group paper identified these capabilities as key parameters for any future space traffic management organization \cite{iaf2026}.

\subsection{Limitations \& Constraints}
There are technical limitations. Objects smaller than roughly $10~cm$ remain largely undetectable. The $1-10~cm$ space debris population is estimated at around 1.2 million from modeling but is not tracked. So most of the threats are invisible to space operators. The identification system for particular space debris responsible for the damage is poor. After a collision, determining which object caused the damage and which state or operator is responsible requires high-quality tracking and database systems. But databases are rarely available, thereby undermining fault-based liability (Section~\ref{sec5p2}). Data-sharing practices and protocols are poor across national, commercial, and military systems. Here, the \texttt{U.S. Space Data Center} and \texttt{ESA's SST Cooperation} only provide partial database support. The ground-based coverage of GEO and MEO is fundamentally angle-limited, which is why the space-based SSA constellations are now being deployed.

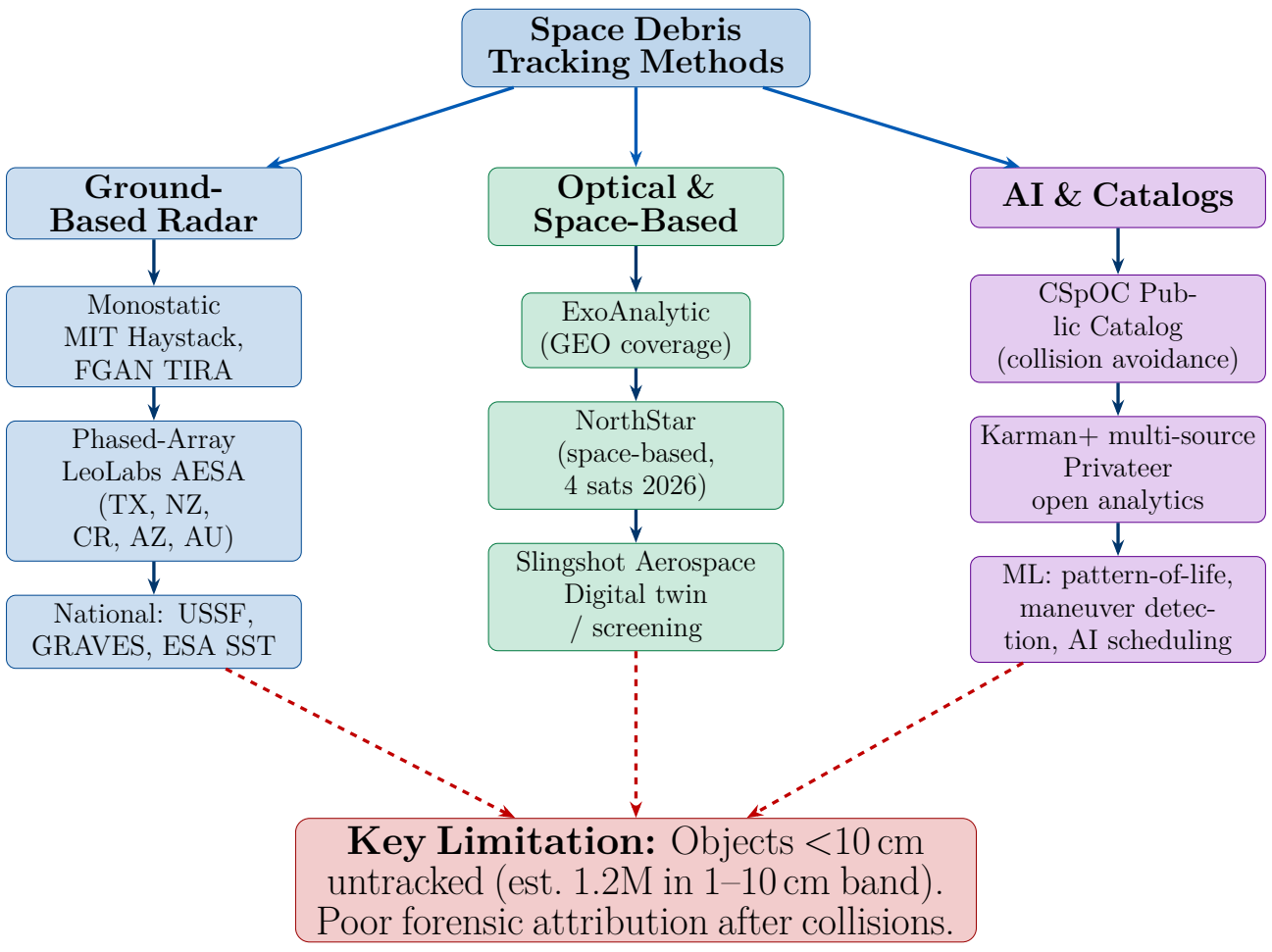
\begin{figure}[htbp]
\centering
\resizebox{0.475\textwidth}{!}{
\begin{tikzpicture}[node distance=0.8cm and 0.6cm]

  \node[blockR=myblue, text width=5cm] (root)
    {\Large \textbf{Space Debris Tracking Methods}};

  \node[smallblock=myblue, below left=1.2cm and 2.4cm of root, text width=4.2cm]
    (radar) {\Large \textbf{Ground-Based Radar}};
  \node[smallblock=mygreen, below=1.2cm of root, text width=4.2cm]
    (optical) {\Large \textbf{Optical \& Space-Based}};
  \node[smallblock=mypurple, below right=1.2cm and 2.4cm of root, text width=4.2cm]
    (ai) {\Large \textbf{AI \& Catalogs}};

  \node[smallblock=myblue, below=0.7cm of radar, text width=4.2cm]
    (r1) {\large Monostatic\\MIT Haystack, FGAN TIRA};
  \node[smallblock=myblue, below=0.5cm of r1, text width=4.2cm]
    (r2) {\large  Phased-Array\\LeoLabs AESA\\(TX, NZ, CR, AZ, AU)};
  \node[smallblock=myblue, below=0.5cm of r2, text width=4.2cm]
    (r3) {\large  National: USSF, GRAVES, ESA SST};

  \node[smallblock=mygreen, below=0.7cm of optical, text width=3.2cm]
    (o1) {\large  ExoAnalytic\\(GEO coverage)};
  \node[smallblock=mygreen, below=0.5cm of o1, text width=4.2cm]
    (o2) {\large  NorthStar\\(space-based, 4 sats 2026)};
  \node[smallblock=mygreen, below=0.5cm of o2, text width=4.2cm]
    (o3) {\large  Slingshot Aerospace\\Digital twin / screening};

  \node[smallblock=mypurple, below=0.7cm of ai, text width=4.2cm]
    (a1) {\large  CSpOC Public Catalog\\(collision avoidance)};
  \node[smallblock=mypurple, below=0.5cm of a1, text width=4.2cm]
    (a2) {\large  Karman+ multi-source\\Privateer open analytics};
  \node[smallblock=mypurple, below=0.5cm of a2, text width=4.2cm]
    (a3) {\large  ML: pattern-of-life,\\maneuver detection, AI scheduling};

  \node[wideblock=myred,
        below=1.0cm of optical,
        yshift=-7.6cm, text width=10cm]
    (lim)
    {\LARGE \textbf{Key Limitation:}
     Objects $<$10\,cm untracked (est.\ 1.2M in 1–10\,cm band).
     Poor forensic attribution after collisions.};

  \foreach \t in {radar, optical, ai}{
    \draw[arrow=myblue] (root) -- (\t);
  }
  \foreach \s/\t in {radar/r1,r1/r2,r2/r3,
                     optical/o1,o1/o2,o2/o3,
                     ai/a1,a1/a2,a2/a3}{
    \draw[arrow=myblue!60!black] (\s) -- (\t);
  }
  \foreach \s in {r3,o3,a3}{
    \draw[arrow=myred, dashed] (\s) -- (lim);
  }

\end{tikzpicture}}
\caption{Space debris tracking methods}
\label{fl3}
\end{figure}

In Figure~\ref{fl3}, the summary of space debris tracking is shown as a schematic diagram.

\section{Active Debris Removal \& End-of-Life Services} \label{sec3}
Currently, three main schools of thought are pursued to tackle the space debris problem. These are in-space collection, remote destruction, and de-orbiting.  

\subsection{Capture \& contact technologies} 
Active debris removal technology is conventionally divided into rigid and flexible, as well as contact and contactless methods. Rigid capture uses robotic arms with heritage from \texttt{Canadarm} and the Space Station remote manipulator. The \texttt{ESA-ClearSpace1} mission applies a four-arm capture concept to a large non-cooperative target \cite{esa2024}, and \texttt{Northrop Grumman's Mission Robotic Vehicle (MRV)}, launched in July 2026, carries twin dexterous arms developed with the \texttt{U.S. Naval Research Laboratory} for inspection, repair and attachment of Mission Extension Pods (MEPs) on GEO satellites \cite{defen2026, 2013ESASP.723E..85W}. Flexible capture includes nets and harpoons demonstrated by the \texttt{RemoveDEBRIS} mission (2018, University of Surrey). The nets enclose irregular tumbling targets at safe standoff distances, while harpoons penetrate and anchor large targets, but this carries fragment-generation risk. \texttt{ESA} is also studying tentacle-based capture for high-angular-momentum targets, and ``space ball''/winch concepts exert retarding torque on tumbling objects. Magnetic capture, demonstrated by \texttt{Astroscale's ELSA-d mission (2021)}, docks a magnetic plate on the servicer (chaser) to a ferromagnetic docking plate on the client (target), simplifying capture for cooperative or prepared targets \cite{sn2021,2026AdSpR..7712265A}. Targets tumble at rates of several degrees per second, so capture requires relative navigation, detumbling, and synchronized contact at centimeter-level precision \cite{abott2025, sn2022}.

\subsection{Contactless \& deorbit-augmentation technologies} 
Contactless methods avoid the collision and fragment-generation risks associated with contact. Laser ablation transfers momentum through photon pressure or vaporization. \texttt{NASA} has studied optimal networks of space-based lasers for debris remediation \cite{WILLIAMSROGERS20255265}. The ion-beam shepherd, developed in Europe, uses a broad, low-divergence ion beam from the servicer (chaser) to deorbit a client (target) without mechanical contact, while conserving the total system momentum. Electrodynamic tethers (e.g., Terminator Tape) generate drag via Lorentz-force interactions with the geomagnetic field, and drag sails passively augment atmospheric drag. China demonstrated the first large-scale deployment of a deorbit sail on a Long March-2D upper stage in 2022 \cite{scio2022}. Propulsive deorbit with chemical or electric propulsion remains the most mature and reliable method for controlled disposal, and is the baseline for GEO graveyard disposal.

\subsection{Rendezvous \& proximity operations}
All ADR and servicing missions are fundamentally rendezvous-and-proximity-operations (RPO) problems against non-cooperative, frequently tumbling targets. The GNC stack comprises absolute orbit determination from ground tracking; relative navigation via vision-based navigation (VBN) using monocular and stereo cameras; LIDAR-based relative ranging; and feature-based identification of the target's structure for capture-point selection. A safe approach requires autonomous collision avoidance during proximity operations, terminal-phase guidance with sub-meter relative accuracy, and robust fault detection. The \texttt{Japan Aerospace Exploration Agency's} \texttt{Commercial Removal of Debris Demonstration (CRD2)} program provides the current example of such a maneuver. \texttt{Astroscale's ADRAS-J satellite (Phase I)} launched in February 2024 and became the first mission ever to approach, inspect, and return close-range imagery of a large real debris object - a Japanese H-IIA upper stage approximately $11~m$ long, $4~m$ in diameter, and $3~tonnes$ in mass - approaching to within roughly $15~m$ before transitioning to a disposal orbit compatible with natural decay \cite{sdc9paper121}. \texttt{Phase II, ADRAS-J2}, will attempt the world's first capture and removal of that same large debris object\cite{sdc9paper121}.

\subsection{Mission case studies}
Table~\ref{op_stat} summarizes the reference missions that define the technology readiness baseline as of 2026. \texttt{ESA's} Zero Debris approach, adopted in 2023–2024, enforces the agency and its member states to effectively cease debris generation in protected orbital regions by 2030 and integrates stricter disposal requirements into procurement — an institutional pull that, together with the FCC's binding five-year deorbit rule (Section~\ref{sec5p2}), converts EOL compliance from guideline to contract \cite{esa2023, fcc2022}.

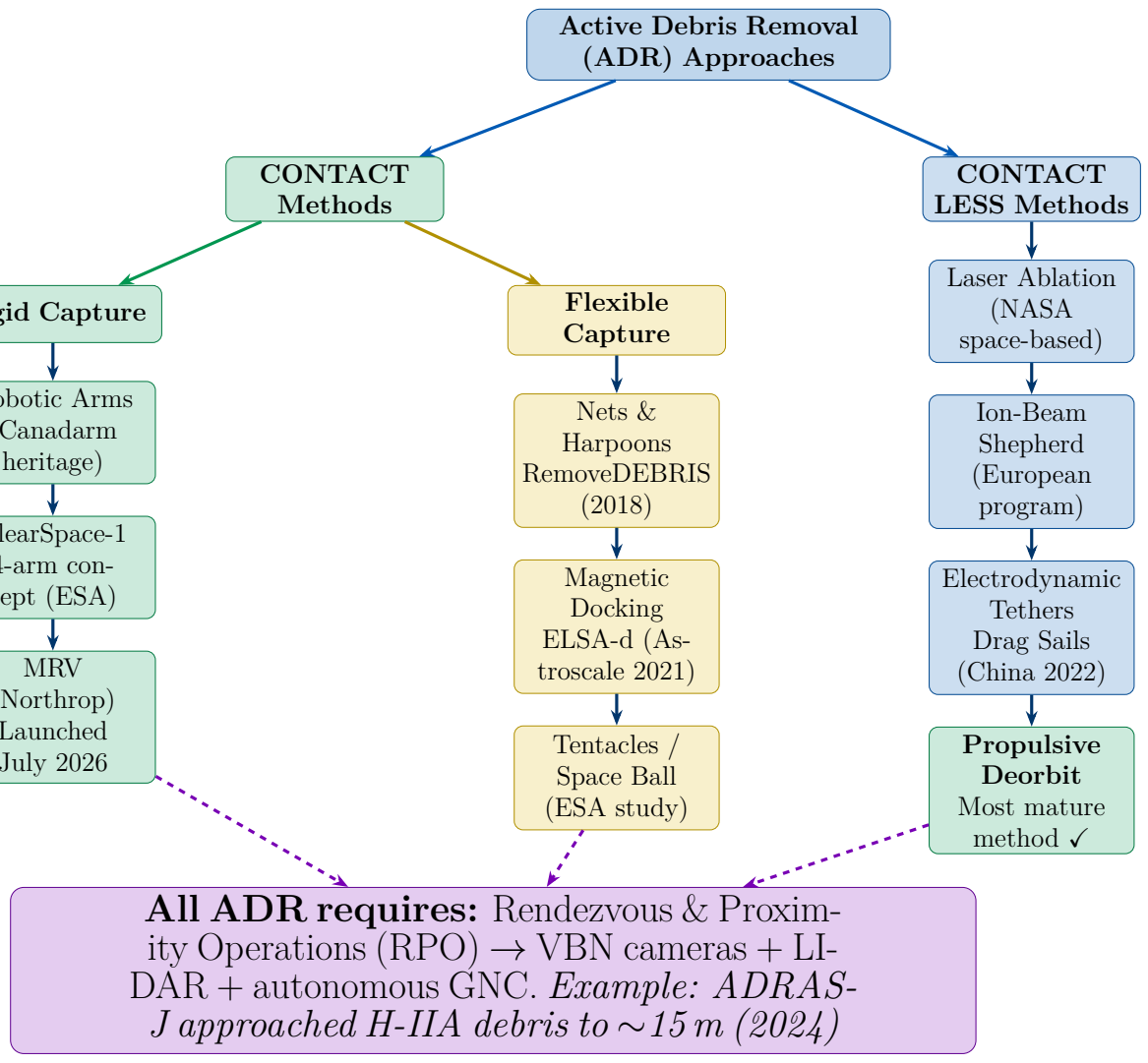
\begin{figure}[htbp]
\centering
%
\resizebox{0.45\textwidth}{!}{
\begin{tikzpicture}[node distance=0.9cm and 0.5cm]

  \node[blockR=myblue, text width=5.5cm] (root)
    {\textbf{Active Debris Removal (ADR) Approaches}};

  \node[smallblock=mygreen, below left=1.2cm and 1.3cm of root]
    (contact) {\textbf{CONTACT Methods}};
  \node[smallblock=myblue, below right=1.2cm and 0.5cm of root]
    (nocontact) {\textbf{CONTACT\\LESS Methods}};

  \node[smallblock=mygreen, below left=1.0cm and 1.0cm of contact]
    (rigid) {\textbf{Rigid Capture}};
  \node[smallblock=myyellow, below right=1.0cm and 1.0cm of contact]
    (flex) {\textbf{Flexible Capture}};

  \node[smallblock=mygreen, below=0.6cm of rigid, text width=3.0cm]
    (rg1) {\large  Robotic Arms\\(Canadarm heritage)};
  \node[smallblock=mygreen, below=0.5cm of rg1, text width=3.0cm]
    (rg2) {\large  ClearSpace-1\\4-arm concept (ESA)};
  \node[smallblock=mygreen, below=0.5cm of rg2, text width=3.0cm]
    (rg3) {\large  MRV (Northrop)\\Launched July 2026};

  \node[smallblock=myyellow, below=0.6cm of flex, text width=3.0cm]
    (fl1) {\large  Nets \& Harpoons\\RemoveDEBRIS (2018)};
  \node[smallblock=myyellow, below=0.5cm of fl1, text width=3.0cm]
    (fl2) {\large Magnetic Docking\\ELSA-d (Astroscale 2021)};
  \node[smallblock=myyellow, below=0.5cm of fl2, text width=3.0cm]
    (fl3) {\large  Tentacles / Space Ball\\(ESA study)};

  \node[smallblock=myblue, below=0.6cm of nocontact, text width=3.0cm]
    (nc1) {\large  Laser Ablation\\(NASA space-based)};
  \node[smallblock=myblue, below=0.5cm of nc1, text width=3.0cm]
    (nc2) {\large  Ion-Beam Shepherd\\(European program)};
  \node[smallblock=myblue, below=0.5cm of nc2, text width=3.0cm]
    (nc3) {\large  Electrodynamic Tethers\\Drag Sails (China 2022)};
  \node[smallblock=mygreen, below=0.5cm of nc3, text width=3.0cm]
    (nc4) {\large  \textbf{Propulsive Deorbit}\\Most mature method $\checkmark$};

  \node[wideblock=mypurple,
        below=10.5cm of contact,
        xshift=2.5cm, text width=15.0cm]
    (gnc)
    {\LARGE \textbf{All ADR requires:} Rendezvous \& Proximity Operations (RPO)
     $\to$ VBN cameras + LIDAR + autonomous GNC.
     \textit{Example: ADRAS-J approached H-IIA debris to $\sim$15\,m (2024)}};

  \draw[arrow=myblue] (root) -- (contact);
  \draw[arrow=myblue] (root) -- (nocontact);
  \draw[arrow=mygreen] (contact) -- (rigid);
  \draw[arrow=myyellow!80!black] (contact) -- (flex);
  \foreach \s/\t in {rigid/rg1,rg1/rg2,rg2/rg3,
                     flex/fl1,fl1/fl2,fl2/fl3,
                     nocontact/nc1,nc1/nc2,nc2/nc3,nc3/nc4}{
    \draw[arrow=myblue!60!black] (\s) -- (\t);
  }
  \foreach \s in {rg3,fl3,nc4}{
    \draw[arrow=mypurple, dashed] (\s) -- (gnc);
  }

\end{tikzpicture}}
\caption{ADR Technology}
\label{fl4}
\end{figure}

Figure~\ref {fl4} summarizes the latest ADR technologies in a diagram. 

\section{The Commercial Landscape} \label{sec4}

\subsection{SSA data as a service}
Space Situational Awareness(SSA) data was the first sustainable product to reach genuine commercialization. The customer base spans satellite operators (collision avoidance and maneuver planning), launch providers (range safety and on-orbit verification), insurers (underwriting and claims forensics), and defense and civil agencies (threat assessment, SDA). Revenue models are subscription-based data-as-a-service (DaaS), practical surveillance contracts, and government contracts. \texttt{LeoLabs}, started with collision protection services for operators, then moved toward the defense sector.
It won a \$20.68 million \texttt{U.S. Space Force} contract for Scout radar services in 2026, after the December 2025 catalog license with the Office of Space Commerce and Space Force, and reported 186 percent growth in U.S. government contract bookings in 2025 \cite{orbin2026, nse2026a}. \texttt{NorthStar's} public-market listing in 2026 marks the entry of SDA into public equity markets \cite{nse2026b}. \texttt{Slingshot, ExoAnalytic, Karman+ and True Anomaly (responsive spacecraft)} complete a layered commercial ecosystem.

\subsection{Remediation, disposal \& in-orbit servicing} 
Active debris removal (ADR) and End-of-Life (EOL) services remain dependent on institutional customers. The \texttt{ClearSpace} has transitioned from the \texttt{ESA-funded ClearSpace-1} to a GEO life-extension line (Phoenix, contracted by ESA in September 2026), moving up the value chain from removal to revenue-generating service \cite{lsa2026}. Space logistics operationalized the leasing model, in which MEV-1/MEV-2 sell mission extension time to GEO operators rather than hardware, and the 2026 MRV/MEP launch extends this to include inspection and repair \cite{defen2026}. \texttt{OrbitFab} is pursuing in-orbit refueling infrastructure (fluid-transfer interfaces and orbital fuel depots), while \texttt{D-Orbit, Starfish Space, Kinetik}, and others target the LEO disposal and orbital transfer vehicle (OTV) segment. \texttt{Novaspace (Euroconsult)} projects roughly \$3 billion in cumulative in-orbit-services revenue over the coming decade \cite{smr2026}.

\subsection{The space insurance market}
The insurance sector in the space business is essential, but it is not yet mature. In 2023-24, the space insurance market produced the largest losses in over two decades, on the order of $\$800$ million from the \texttt{Viasat-3} and \texttt{Inmarsat-6 F2} in-orbit failures before stabilizing in 2025 \cite{wtw2025}. The loss of \texttt{SpainSat NG2} in late 2025, with claims estimated at approximately \$400 million, also weighed on 2026 underwriting results \cite{bd2026}. The insurance rate increase has been severe, with launch premiums rising sharply after the 2023 losses \cite{wtw2025}. In China, a commercial space co-insurance consortium launched in March 2025 provided \$1.5 billion of insurance cover in its first year \cite{inasia2026}.

\subsection{Business models}
Table~\ref{tab_biz} classifies the business models that have emerged. The customers are divided into two classes. On the one hand, governments and defense agencies are the primary customers for SSA/SDA and ADR, while, on the other hand, commercial operators contribute to the scaling demand for data services and, increasingly, EOL compliance services driven by regulation. The insurers are at the junction; they fear price-collision risk, and their underwriting requirements (SSA data usage, disposal plans) further propagate sustainability requirements into mission design and cost.

\section{Insurance \& Liability Architecture}\label{sec5} 

\subsection{The insurance gap as an incentive gap} 
The space insurance market exhibits all the classical failure modes of an emerging catastrophe line. The small premium base has limited exposure, and it shows a lack of communication between the underwriter and the operator, leading to moral hazard in maneuvering and debris-disposal behavior. The risk of correlated loss is the main threat in this sector. Third-party liability (TPL) coverage is small relative to the asset values.
Due to the lack of fault attribution, the liability setup is very poor. The result is an insurance gap that appears as the remediation gap. The risk is priced too low to cover cleanup, and no one is responsible for the damage. The community's responsibility keeps deteriorating. The financial innovation agenda should focus on insurance-backed disposal warranties, collision insurance, disposal deposits, and performance bonds, which are precisely attempts to convert these gaps into tradable instruments \cite{wef2024}.

\subsection{The liability architecture} \label{sec5p2}
The legal framework comprises the 1967 Outer Space Treaty (OST) and the 1972 Convention on International Liability for Damage Caused by Space Objects. The OST establishes that States Parties, who bear international responsibility for national activities in outer space (Article VI), are internationally liable for damage caused by their space objects (Article VII), and shall conduct activities with ``due regard'' to the corresponding interests of other States and avoid harmful contamination (Article IX). The Liability Convention operationalizes Article II, which imposes absolute liability on the launching State for damage caused on the surface of the Earth or to aircraft in flight; and Article III, which imposes fault-based liability for damage caused elsewhere (i.e., in space) to another launching State's space object or persons or property on board; and Article IV, which provides joint and several liability with apportionment between jointly launching States \cite{un1972, unoosa1972, iaf2026}.

The remediation economics depends upon liability running through States, not operators. An operator's exposure is mediated by its licensing State, which typically indemnifies the liability as a condition of license. The Convention compensates damage but imposes no obligation to remove debris. There is no international legal mandate to remediate, and no liability attaches to the mere existence of a derelict object. The fault-based liability in space is nearly inoperative in practice because attribution is technically immature (Section 2). Without reliable pre-collision custody, the ``fault'' of a specific object and operator cannot be proven, leaving the space operators' principal deterrent channel idle.

National authorities show reluctance in how states communicate these obligations into licensing and financing. In the United States, the Commercial Space Launch Act requires launch licensees to obtain TPL insurance or pay for financial responsibility around \$500 million or the maximum available on the world market. This comes with federal indemnification of covered third-party claims above that amount up to \$1.5 billion (as adjusted for inflation) (14 CFR 440.5, 440.9). The FCC's orbital debris rules, adopted in the Second Report and Order (FCC 22-74) and codified at 47 CFR 25.283(e), impose a binding requirement that FCC-licensed satellites ending their mission in or passing through LEO below 2,000 km and planning uncontrolled re-entry complete disposal no later than five years after end of mission, which is effective since  September 2024. This rule replaced the voluntary 25-year guideline and, for the first time, created a binding regulatory requirement for disposal services \cite{fcc2022, slj2026}. The United Kingdom's Outer Space Act 1986 and Space Industry Act 2018 impose a \euro{60 million} TPL insurance/indemnity limit for standard missions. From February 2026, all SIA 2018 licenses include mandatory liability limits, ending the unlimited-liability exposure identified by the sector \cite{ukgov2026}. France's Law on Space Operations (LOS 2008-518) makes the operator strictly responsible for third-party damage (fault of the victim excepted) and provides that the State substitutes for the operator one year after the fulfillment of licensing obligations, indemnifying claims above the operator's prescribed limit \cite{legifr2008}. China licenses space launches and satellites through its civil launch and satellite radio licensing authorities, and its 2025 space insurance consortium indicates a deliberate build-out of domestic underwriting capacity \cite{inasia2026}.

International soft-law and standards complete this framework. \texttt{The United Nations Committee on the Peaceful Uses of Outer Space (UNCOPUOS)} Guidelines for the Long-Term Sustainability of Outer Space Activities (21 guidelines adopted in 2019) recommend that States adopt debris mitigation and remediation practices \cite{unoosa2019}. The IADC Space Debris Mitigation Guidelines (2007, updated 2025) provide the technical baseline for disposal \cite{unoosa2026}. The ISO 24113:2023 (4th edition) converts mitigation requirements into an auditable engineering standard \cite{iso2023}. The ITU's Radio Regulations, WRC-23 decisions, and Resolutions at ITU-R 74 link spectrum, for orbital-resource authorization to de-orbit and disposal strategy, extend the sustainability conditions into spectrum licensing 
\cite{itu2023, itu2024}.

\section{Commercial Space Sustainability Market}\label{sec6} 
\subsection{The coupled incentive architecture} 
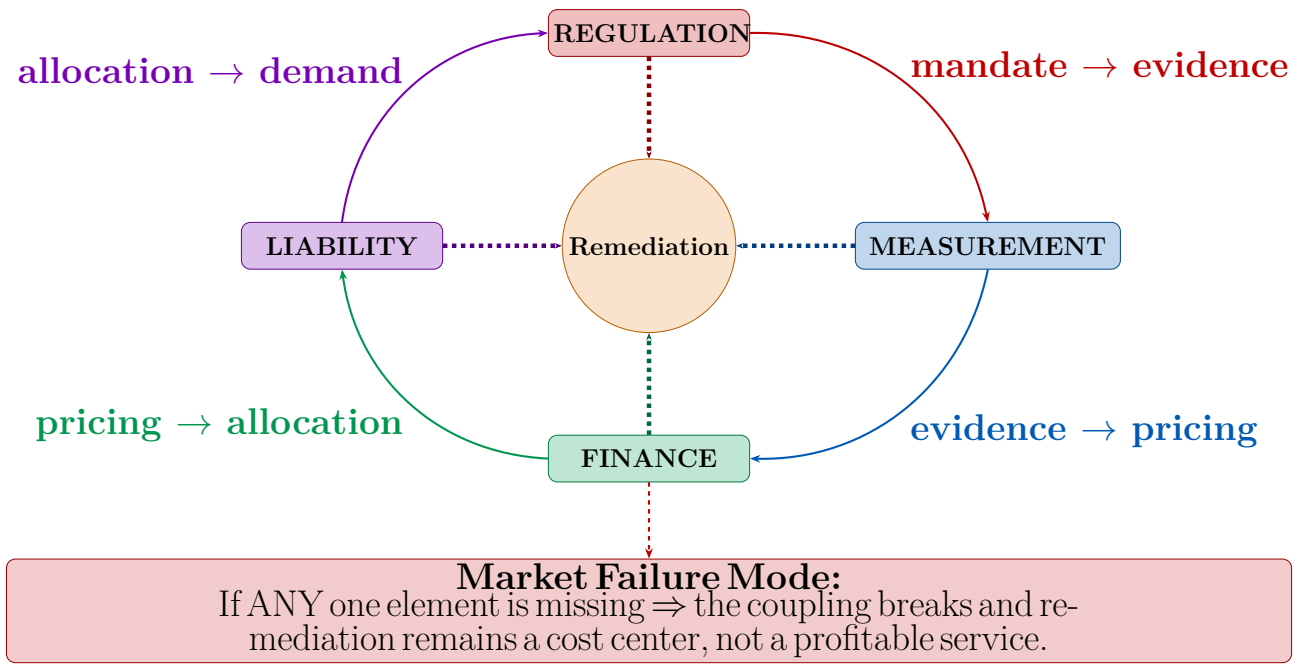
\begin{figure}[htbp]
\centering

\resizebox{0.475\textwidth}{!}{
\begin{tikzpicture}[node distance=2.0cm]

  \node[circle, draw=myorange!80!black, fill=myorange!20,
        text width=3.8cm, align=center, font=\LARGE\bfseries,
        minimum size=3.8cm] (center)
    {Remediation};

  \node[blockR=myred, above=2.4cm of center, text width=4.5cm]
    (reg)
    {\LARGE \textbf{REGULATION}}; 

  \node[blockR=myblue, right=2.8cm of center, text width=6cm]
    (meas)
    {\LARGE \textbf{MEASUREMENT}};

  \node[blockR=mygreen, below=2.4cm of center, text width=4.5cm]
    (fin)
    {\LARGE  \textbf{FINANCE}}; 

  \node[blockR=mypurple, left=2.8cm of center, text width=4.5cm]
    (liab)
    {\LARGE  \textbf{LIABILITY}}; 

  \draw[arrow=myred, bend left=40]
    (reg.east) to node[above right, font=\Huge\color{myred}]
    {\bfseries mandate $\to$ evidence} (meas.north);
  \draw[arrow=myblue, bend left=40]
    (meas.south) to node[below right, font=\Huge\color{myblue}]
    {\bfseries evidence $\to$ pricing} (fin.east);
  \draw[arrow=mygreen, bend left=40]
    (fin.west) to node[below left, font=\Huge\color{mygreen}]
    {\bfseries pricing $\to$ allocation} (liab.south);
  \draw[arrow=mypurple, bend left=40]
    (liab.north) to node[above left, font=\Huge\color{mypurple}]
    {\bfseries allocation $\to$ demand} (reg.west);

  \foreach \s/\c in {reg/myred, meas/myblue,
                     fin/mygreen, liab/mypurple}{
    \draw[-{Stealth[length=5pt]}, \c!70!black,
          line width=3pt, dashed] (\s) -- (center);
  }

  \node[wideblock=myred, below=1.8cm of fin, text width=30cm]
    (fail)
    {\Huge \textbf{Market Failure Mode:}\\
     If ANY one element is missing $\Rightarrow$
     the coupling breaks and remediation remains a cost center,
     not a profitable service.};

  \draw[arrow=myred, dashed] (fin.south) -- (fail.north);

\end{tikzpicture}}
\caption{The Coupled Incentive Architecture}
\label{fl5}
\end{figure}

Sustainability becomes a business only when the following four elements are coupled. They are: a binding disposal mandate (demand), commercial-grade SSA with attribution capability (evidence), insurance and financial instruments that price and transfer EOL and collision risk (finance), and liability rules that allocate fault and cost (law). Each element is interdependent with the others. 
Observation \& tracking without any mandate finds no paying customer. The policy/regulation that cannot punish/prosecute a defaulter is just paperwork. Liability rules that cannot provide any remediation for the risk are not useful. We have watched markets fail whenever any one element was missing. The commercial models in Section~\ref{sec4} already instantiate the coupling at the firm level. \texttt{Astroscale} sells both inspection (\texttt{ADRAS-J}) and removal (\texttt{ADRAS-J2}) to an agency that simultaneously procures sustainability.
\texttt{LeoLabs} sells the data that insurers and regulators need. The insurers depend on planned disposal, but the system-level interaction remains incomplete. Figure~\ref{fl5} shows the diagram of the coupled incentive architecture.

\subsection{Technical recommendations} 
Standard ADR interfaces and design-for-demise (D4D) procedures in international standards are required so that any servicer (chaser) can capture any compliant target. Using space-based SSA constellations along with data models, the tracked population of space debris can be extended to sizes smaller than 1-10 cm. This will develop forensic attribution standards (pre-collision custody, conjunction data provenance) to enable fault-based liability, as well as a common disposal telemetry database accessible to all players.

\subsection{Commercial \& financial recommendations} 
To execute the plan, we need to start the initiative and then commercialize procurement, as \texttt{JAXA} and \texttt{ESA} have done with explicitly contracted downstream licensing of mission-proven technology.
This will make the end-of-life deposits refundable upon verified disposal, which will be administered through national licensing authorities.
This will develop insurance-linked disposal warranties and parametric collision products with public seed capacity and extend sustainability ratings (SSR) into procurement preferences and insurance pricing so that verified sustainability becomes a market premium rather than a cost \cite{ssr2025}.

\subsection{Policy recommendations \& the HKSAR perspective}
Policies need to be developed to harmonize national licensing conditions with the IADC baseline and ISO 24113. Under this, all negotiations on attribution and apportionment should be conducted in accordance with the \texttt{UNCOPUOS} and \texttt{ITU} guidelines. This will help mitigate disposal-related issues and the spectrum fees associated with any misuse. 

Hong Kong SAR occupies a distinctive position in this scenario as a global insurance and financial center with a deep heritage in marine insurance and arbitration. With legal talent active in space law and dispute resolution, Hong Kong SAR could host an orbital insurance and sustainability finance hub, providing the broking, underwriting, structuring, and dispute resolution infrastructure that the instruments in Section~\ref{sec5} require, thereby complementing its growing role in NewSpace science and industry. In the 15th Five-Year Plan (2026-30), the Hon. Chief Executive of HKSAR envisioned HKSAR as an emerging hub for the global newspace economy \cite{hk5, cd202609}. 

\subsection{NGOs -  the necessity \& significance}\label{sec6p5}
The incentive architecture discussed above presumes an institutional layer that neither States nor markets currently supply. International space governance is state-centric. The \texttt{Outer Space Treaty} and the Liability Convention bind the State Parties. The \texttt{UNCOPUOS} operates by consensus, and national regulators move at the pace of domestic legislative cycles. Remediation, in contrast, is a fast-moving, industry-driven problem. Between the two lies a coordination gap that non-governmental organizations (NGOs) are well-suited to fill, which is precisely the rationale for initiatives such as \texttt{S3+1}, a non-profit Hong Kong SAR-based but inherently international organization with a deeply respected board\footnote{\url{https://s3plus1.org/}}.

An NGO provides neutral trust infrastructure. Attribution standards, data-sharing governance, and the pricing of collision risk require institutions that neither defense agencies (whose SSA data are dual-use and sovereignty-bound) nor commercial firms (whose incentives are profit-bound) can credibly provide. An independent, mission-driven NGO can convene operators, insurers, scientists, and civil society across jurisdictions and mandate horizons, lowering the transaction costs of the coupled incentive architecture. NGOs are also the historic incubators of soft law. The IADC mitigation guidelines, the Space Sustainability Rating, and successive \texttt{United Nations Institute for Disarmament Research (UNIDIR)} and \texttt{World Economic Forum (WEF)} financial-mechanism dialogues all began as informal, multi-stakeholder efforts before strengthening into regulations. An NGO can create disposal deposits, orbit-use fees, and sustainability ratings ahead of the regulatory cycle, de-risking them for later codification. Organizations such as \texttt{S3+1} deliberately bridge scientific, technological, economic, and societal perspectives, a mandate that \texttt{Science, Technology, and Economic Policy(STEP)} itself serves. This gives sustainability a forum in which its economic value can be articulated, measured, and contested before it is priced. The weakest-link problem also demands an accountability function in which an independent voice upholds transparency benchmarks, monitors compliance with disposal, and represents the orbital commons in licensing and liability proceedings. Additionally, NGOs can conduct capacity-building for emerging space actors, thereby diffusing norms and a stewardship culture beyond the founding members of the orbital economy. 
It does not set the price of remediation only. It makes the price worth trusting.

\subsection{Stakeholder collaboration \& operational implementation} 
The governance architecture implies collaboration across four classes of actors (States and regulators, intergovernmental organizations, non-governmental institutions and commercial firms), linked together by five areas of collaboration, i.e., (1) the contractual agreements (licensing and procurement), (2) the data sharing (SSA sharing and commercial catalog licensing), (3) the trust (rating and accountability), (4) the financial collaborations (insurance, deposits and securities) and (5) the dispute-resolution policies (arbitration and liability apportionment). Commercialization is not the autonomous growth of a market; it is the conversion of regulatory obligations into transaction infrastructures that are priceable, transferable, and verifiable. That conversion works under the following conditions:\\

(1) Regulation-driven demand comes first. Binding disposal mandates, such as the FCC five-year rule (47 CFR 25.283(e)), to convert an externality into a compliance obligation, generate demand for disposal-as-a-service supplied by ADR and end-of-life vendors on per-spacecraft or launch-bundled terms. Compliance evidence comprises a disposal contract and telemetry attestation. The binding constraint is jurisdictional, as the rule applies only to FCC-licensed satellites and invites registry shopping. \\
(2) Anchor procurement and commercial replication follow as the governments de-risk first-of-kind technologies through first-mover contracts (\texttt{JAXA CRD2 Phases I–II; ESA ClearSpace-1}), then extend the service for commercial customers (\texttt{Astroscale–Eutelsat; ClearSpace Phoenix}). Implementation depends on fixed-price milestone payments and data rights agreements, with export control authorities (ITAR/EAR) as the principal regulators. \\
(3) Insurance-linked transmission works through underwriting. The underwriters' conditions depend on the use of SSA data and on disposal plans, translating the sustainability obligation into billed amounts, while the collision coverage pays on event triggers. For this, reinsurance structures and standard policy regulations are required with underwriting capacity.
The annual premiums of roughly US\$500–600 million, together with forensic evidence chains, constitute the binding constraints. \\
(4) Disposal of deposits and producer-responsibility procedures take the form of upfront end-of-life deposits, which will be refunded upon verified disposal, creating a ring-fenced fund that gives the ADR industry a predictable cash flow. The implementation requires escrow governance, verified telemetry, and a harmonized valuation formula, and no State has yet to legislate the process. \\
(5) Securitization, through \texttt{Space Debris Retrieval Insurance Bonds (SPADRIBs)}\footnote{\url{https://www.longfinance.net/publications/professional-articles/space-debris-through-the-telescope-of-a-financial-analyst/}}, converts disposal obligations into bonds serviced by disposal fees and avoided-collision value. This is under discussion in the London market and requires standardized asset pools, rating infrastructure, and cash-flow modeling. This is in a pre-commercial stage now. \\
(6) Rating premia arise when sustainability ratings (e.g., the Space Sustainability Rating, operated by \texttt{EPFL} and \texttt{OKAPI}) feed procurement scoring, ESG-linked lending and insurance pricing, making verified sustainability a market premium through certification–disclosure–audit loops. \\
(7) Fee-funded public goods are the last piece. The orbit and spectrum fees on a polluter-pays basis would capitalize a governance fund for SDA data sharing and ADR, while statutory indemnification (the US government-backed institutions to US\$1.5 billion; the UK Space department beyond \euro{(60 million)} regularizes systemic third-party liability risk. International coordination is required to prevent capital flight.\\

Across all the mechanisms, the division of labor should be stable. The States create obligations and the lion's share of the demand; intergovernmental organizations harmonize baselines; NGOs supply neutral trust infrastructure, create soft law and exercise accountability (Section~\ref{sec6p5}); and firms/companies supply technology and close the loop. The critical failure modes, such as jurisdictional arbitrage, export controls, underwriting capacity, and attribution evidence, are precisely the gaps that coordination and standard-setting must finalize before these mechanisms scale up.

\subsection{The future industry landscape}
Between 2026 and 2040, three factors will reshape the orbital economy as regulators tighten disposal rules, orbital risk is priced and packaged as a financial product, and on-orbit operations get industrialized. The demand for space launches alone will not demonstrate the value, but the interaction of those three factors will.

Binding disposal mandates (the FCC five-year rule, 47 CFR 25.283(e); UK, French and EU policies) are proliferating but remain legally uneven. Genuine sustainability requires convergence on a common baseline through IADC, ISO 24113 and UNCOPUOS, because only universal obligations can convert remediation from voluntary demonstration into a priced service category. The scope of dispute resolution is itself a commercial niche due to a lack of communication among the States. 
As forensic SSA matures, collision and disposal risk migrates from a balance-sheet contingency to a tradable business with insurance-linked warranties, disposal deposits, SPADRIB-type bonds, and sustainability ratings. The space-insurance premium base (roughly US\$500–600 million per year) is small relative to the liability and remediation scope it must eventually cover. This indicates a structural growth direction tied to data sharing rather than to launch cadence. 
ADR is converging with in-orbit servicing, refueling, towing, and recycling into an orbital logistics economy (\texttt{MEV, MRV, ClearSpace Phoenix, CRD2, SJ-21}), where remediation becomes a route rather than a mission, and disposal becomes a regulated logistics service with recurring revenue. This opportunity follows directly as below:
\begin{enumerate}
\item Reliable SSA and collision-risk assessment -- sold to underwriters, operators, and regulators on recurring contracts (\texttt{LeoLabs, NorthStar, ExoAnalytic}). 

\item Disposal-as-a-service and compliance attestation -- regulation-driven, per-spacecraft and launch-bundled. 

\item In-orbit servicing and life-extension -- the first segment is implemented very recently (\texttt{MEV–Intelsat 901; ClearSpace Phoenix}). 

\item Remediation funds, deposit escrow and orbital asset management -- financial intermediation that secures deposits and structures SPADRIBs, earning stewardship margins rather than launch margins. 

\item Sustainability rating, disclosure and audit -- a certified economy-driven procurement scoring system for ESG-linked finance and insurance pricing (SSR). 

\item Parametric collision coverage and reinsurance structures -- event-triggered policies that decouple claims from slow attribution processes. 

\item On-orbit recycling and manufacturing -- closing the loop from removal to circularity, with higher value capture on a longer horizon. 

\item Capacity-building and emerging-market access norms -- diffusion and training for new entrants, complementing the NGOs.
\end{enumerate}

There have to be universal obligations, with no safe haven to run to. Any incident must be verifiable and backed by reliable SSA and telemetry data. The financial policies have to be pre-funded and transferable. Attribution and apportionment have to be clear enough for pricing. And someone legitimate has to keep the transaction costs low. On present trajectories, the commercial deadline will likely be reached in the early 2030s, when collision-risk pricing, disposal fees, and servicing revenues converge; until then, government procurement and statutory indemnification should remain the bridge.


\subsection{The Hong Kong SAR advantage}

HKSAR's geographic and socio-economic advantages are not a coincidence. It is a common-law jurisdiction with one of the world's largest insurance and reinsurance markets, a marine insurance heritage that maps directly onto orbital risk transfer, and an internationally preeminent arbitration seat (\texttt{HKIAC}), capable of resolving the attribution and apportionment disputes that Section~\ref{sec5} anticipates. Its financial infrastructure, with free capital movement, large capital markets, and an expanding sustainable-finance ecosystem, can host the deposit escrows, SPADRIB-type bonds, parametric products, and ESG-linked structures that the framework requires, without the licensing conflicts of less-open markets. Under the ``one country, two systems'' framework, HKSAR occupies a unique pivotal position that connects mainland China regulatory, industrial, and market demand (including the insurance consortium and the SJ-21 demonstration) with international treaty policies, Western insurance capacity, and global capital, while preserving neutrality through juridical distance. It has a growing space-science and space-law talent base — including the \texttt{HKU Laboratory for Space Research}\footnote{\url{https://www.lsr.hku.hk/}} — capable of supplying interdisciplinary expertise across technical, legal, actuarial, and financial aspects that the framework relies on.

\texttt{S3+1} is the institutional embodiment of this HKSAR-centric initiative. It is established to advance space sustainability through an interdisciplinary, neutral platform. The \texttt{S3+1} is intended to operate from a Hong Kong SAR base and to perform the functions of non-governmental institutions as convening operators, insurers, regulators, and civil society; incubating soft-law policies (disposal deposits, orbit-use fees, sustainability-rating standards) ahead of the regulatory cycle by maintaining transparency benchmarks and exercising accountability on the weakest-link problem, and building capacity for emerging space actors. Within the overall commercial solution, \texttt{S3+1} is not a competitor to markets or states but rather the trust infrastructure that lowers their transaction costs by converting a coordination gap into a profitable framework. The combination of Hong Kong SAR's institutional advantages with a neutral, locally based non-profit turns the policy architecture into an operating reality in which government demand, market capital, and NGO trust converge within a single umbrella.

The implementation will come through stages rather than suddenly. The pilot working groups come first, such as an orbital-insurance task force, parametric-product demonstrations, and a sustainability-rating authority. Then standards incubate, and the market infrastructure will build up. The scale-up stage comes last by linking HKSAR's platform to mainland consortium capacity and international markets. The principal risk is hollowing out of infrastructure without liquidity. The staged approach, documented in real underwriting and disposal of contracts rather than declarations, is needed to avoid the risk.

\section{Space Debris Recycling} \label{sec7}
The fastest route to orbital sustainability is not to minimize a cost but to maximize an asset. The waste-to-energy (WtE) transition in mainland China is a similar controlled experiment in which that inversion occurs.

\subsection{Isomorphism between MSW and space debris} 
Thirty years ago, the municipal solid waste (MSW) problem in China was similar to today's orbital debris issue. The following parallel features define the similarity. It was an unpriced negative externality where the cost fell on neighbors and successors, not on those who produced it, exactly the structure of space debris generation. The disposal was managed through landfill siting, sorting mandates, subsidies, and repeated policy improvisation, with no market price and therefore no supply chain. Incineration was technically mature but economically prohibitive, so the binding constraint was unit cost. There was no party to charge and no party liable to pay. The waste was an object to be hidden, not a commodity to be contested. All of these have a direct orbital counterpart, which is precisely why the comparison is informative.

\subsection{The Chinese Waste-to-Energy experiment} 
\begin{figure}[htbp]
\centering
\resizebox{0.35\textwidth}{!}{
\begin{tikzpicture}[node distance=0.95cm]

  \node[titlebox=myblue, text width=5.5cm] (h1)
    {\Large Orbital Debris (Today $\to$ 2030s)};
  \node[titlebox=mygreen, right=1.2cm of h1, text width=5.5cm] (h2)
    {\Large China WtE Analogy (1990s $\to$ 2023)};

  \node[smallblock=myred, below=0.7cm of h1, text width=5.5cm] (d1)
    {\Large \textbf{Unpriced Externality}\\
      Operators benefit; collision risk socialised};
  \node[smallblock=myred, below=0.7cm of h2, text width=5.5cm] (w1)
    {\Large \textbf{Municipal Solid Waste (1990s)}\\
     Unpriced; cost fell on neighbours};

  \node[smallblock=myorange, below=1.5cm of d1, text width=5.5cm] (d2)
    {\Large \textbf{Institutional Stack Activated}\\
     $\bullet$ FCC 5-yr disposal mandate\\
     $\bullet$ Licensed SSA catalog\\
     $\bullet$ Disposal deposits (EOL)\\
     $\bullet$ Insurance + SPADRIBs\\
     $\bullet$ S3+1 NGO trust layer};
  \node[smallblock=myorange, below=0.7cm of w1, text width=5.5cm] (w2)
    {\Large \textbf{Policy Stack Activated}\\
     $\bullet$ Sorting mandates\\
     $\bullet$ Feed-in tariff (0.65 CNY/kWh)\\
     $\bullet$ Per-tonne treatment fees\\
     $\bullet$ Securitization of cash flows\\
     $\bullet$ Standards body};

  \node[smallblock=myyellow, below=0.7cm of d2, text width=5.5cm] (d3)
    {\Large \textbf{Cost Curve Convergence}\\
     Deorbit: \$60k/kg $\to$ convergence\\
     AI-SSA cuts catalog cost\\
     Serial production of tug vehicles};
  \node[smallblock=myyellow, below=0.7cm of w2, text width=5.5cm] (w3)
    {\Large \textbf{Scale \& Standardisation}\\
    14 Mt (2010) $\to$ 327 Mt/yr (2023)\\
     975 plants operating\\
     Construction costs fell sharply};

  \node[smallblock=mygreen, below=0.7cm of d3, text width=5.5cm] (d4)
    {\Large \textbf{Debris $\Rightarrow$ RESOURCE}\\
      $\sim$\$600B reuse potential\\
     Break-even: early 2030s\\
     Orbital logistics economy};
  \node[smallblock=mygreen, below=0.7cm of w3, text width=5.5cm] (w4)
    {\Large \textbf{Waste $\Rightarrow$ COMMODITY}\\
     $>$83\% MSW incinerated (2023)\\
     Refuse competed for, not hidden\\
     Decade-long transition};

  \foreach \s/\t in {d1/d2, d2/d3, d3/d4}{
    \draw[arrow=myblue] (\s) -- (\t);
  }
  \foreach \s/\t in {w1/w2, w2/w3, w3/w4}{
    \draw[arrow=mygreen] (\s) -- (\t);
  }

  \node[wideblock=mypurple,
        below=2.5cm of d4,
        xshift=3.35cm,
        text width=12cm] (lesson)
    {\Large \textbf{Key Lesson:}
     A coupled incentive stack (mandate + measurement + finance + liability)
     drives unit cost down until a liability becomes an asset.
     On the terrestrial WtE record, a \textit{decadal} transition is within reach.};


\end{tikzpicture}}
\caption{From Liability to Asset — WtE Analogy}
\label{fl6}
\end{figure}
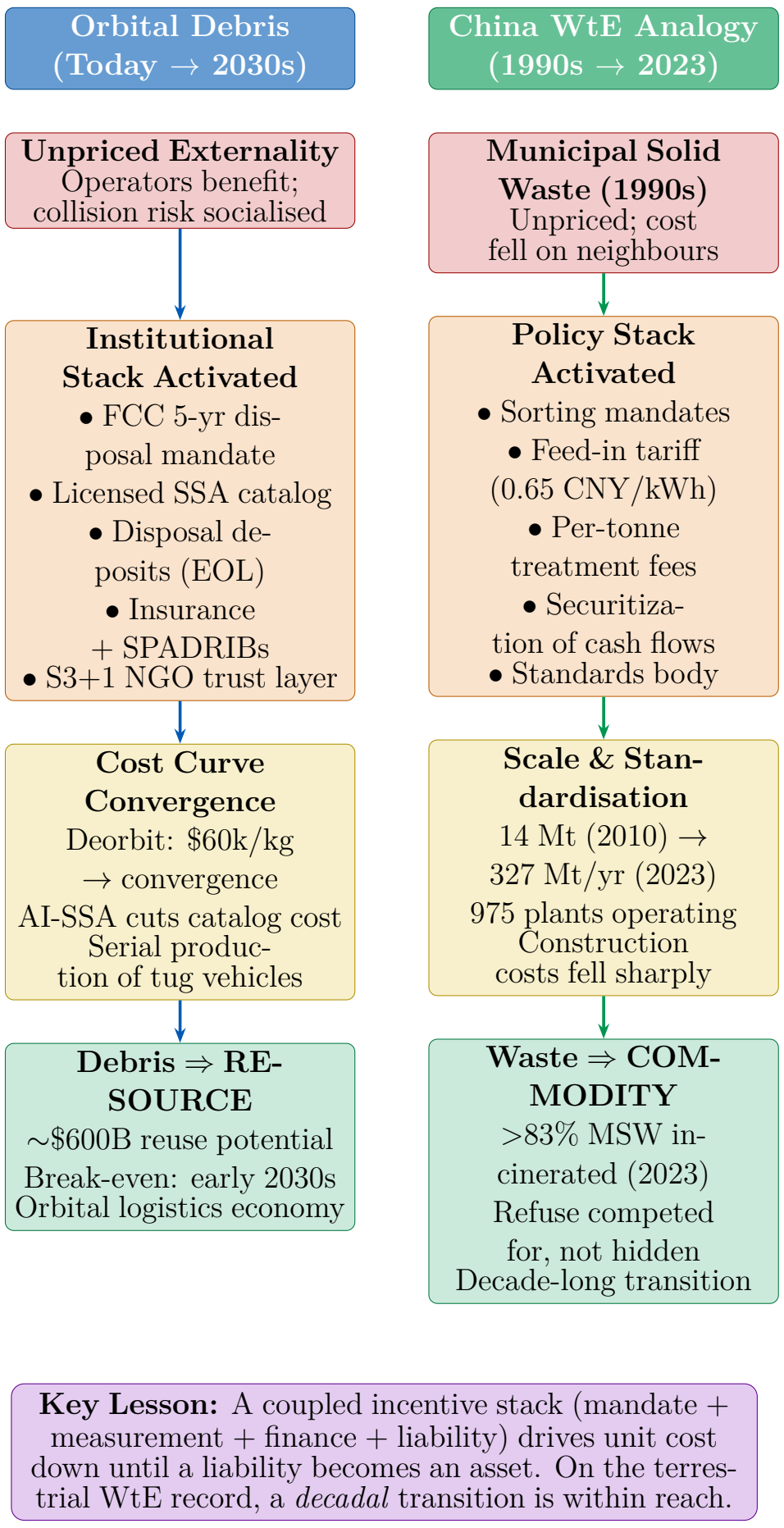

The reversal was obtained with strategies. A stable revenue floor came first with the benchmark feed-in tariff of 0.65 CNY/kWh, making electricity sales remunerative \cite{shen2024}, supplemented by per-tonne treatment fees paid by local governments \cite{wb2010}, although the subsidy phase-downs later reduced the guaranteed revenue as capacity expanded \cite{iea2022}. Scale and standardization then took over as the annual capacity expanded from 14 million tonnes in 2010 to 327 million tonnes in 2023, with 975 plants operating by year-end \cite{cui2026}, and construction costs fell sharply. On the demand side, the loop became self-reinforcing, where electricity and treatment revenues made MSW into positive-value inputs, so incineration came to process over 83 percent of MSW in China by 2023 \cite{cui2026}. As capacity raced ahead of supply, provincial projections indicate persistent overcapacity \cite{shapiro2020}. Within roughly a decade, MSW flipped from a liability to be disposed of into a resource to be competed for. The flip was a cost-curve event, not a technology event: per-tonne electricity output averaged approximately 361 kWh in 2023 and is projected to rise to 497--502 kWh by 2035 as feedstock improves \cite{cui2026}. 
Figure~\ref{fl6} shows the block diagram of the MSW and space debris analogy in the WtE context.

\subsection{Literature review: from externality to asset} 
The existing literature clusters into three groups. One treats debris as pollution and remediation as a cost problem, which are formal economic models of earth-orbit pollution\cite{adilov2015} and of the market for satellites under orbital congestion \cite{bongers2023} internalize the externality through pricing and take-back instruments; Ref.~\cite{rao2020} show that an internationally harmonized orbital-use fee could be more than quadrupled in the long-run for the satellite industry, and added a caution that as the removal lowers congestion, it can raise launch incentives, so remediation without pricing can be welfare-reducing. NASA's cost–benefit capability for mitigating, tracking, and remediating debris \cite{nasa2024} and long-run optimal ADR modeling \cite{bongers2025} belong to the same cost-side tradition, as does the common-pool-resources framing of Ref.~\cite{weeden2012}. 
A second body treats debris as a resource. DARPA's Phoenix program stated the inversion bluntly in 2011 — ``if this program is successful, space debris becomes space resource'' \cite{ssm2011}. An IAC-22 paper argues explicitly for shifting from treating debris as ``a necessary cost that must be minimized'' to ``an untapped space resource and economic opportunity to be maximized'' \cite{iaf2022}. Quantitative research now values orbital debris as a material feedstock for in-space manufacturing, with material-flow analysis extending to 2050 \cite{lenkart2026}. Another circular ``space waste recycling'' architecture \cite{iaf2025} and circular-economy analysis estimated the net reuse potential of on-orbit material at a median of roughly \$600 billion \cite{leonard2023}. 
The third body addresses governance, where polycentric, non-hierarchical arrangements for the orbital commons \cite{morin2021} complement the treaty, rating, and standard material, already reviewed earlier in the article. 
The resource-side literature is conceptual and engineering-led, focused on recycling value and in-space manufacturing, while the economics literature prices debris as a cost. Neither uses a complete, overall industry transition — regulation, tariffs, scale, demand pull, and feedstock competition — as an institutional blueprint, nor does either integrate resource value into the insurance–liability–deposit stack developed in Sections~\ref{sec5} and \ref{sec6}.

\subsection{The realization mechanism}
The debris becomes a resource when the unit cost of rendezvous, capture, processing, and disposal falls below the revenue generated by disposal fees, servicing contracts, avoided-collision value, and recycled feedstock. The record shows both sides of this inequality moving in the same direction as costs fell with scale and standardization, while revenues rose with the tariff and treatment-fee structure. In orbit, the cost side has not yet converged. The estimation for controlled re-entry ranges from \$4,000 to \$60,000 per kilogram and for uncontrolled re-entry from \$3,000 to \$40,000 per kilogram \cite{asaad2026}. A difference of more than an order of magnitude indicates that the technology and market are still before their scale inflection. This is in sharp contrast to the converged cost curve of MSW incineration. The convergence levers are identifiable. There are standardized servicer classes and passive docking interfaces; AI-driven SSA that cuts catalog and conjunction cost; serial production of tug and capture vehicles; and in-space recycling that converts disposal cost into feedstock value. In the WtE context, the incentive framework in Sections~\ref{sec6} and \ref{sec7} maps onto the same machinery. The five-year disposal mandate is the sorting mandate. The licensed commercial SSA catalog is the weighbridge that measures and settles the material flow. The disposal deposit works as the treatment fee, paid in advance. Insurance and SPADRIB pool securitize that flow. The \texttt{S3+1} trust environment develops the tariff and rating standards, and industrial upscaling will close the loop. The WtE record bounds the plausible timescale, and once the stack is coupled, then a significant transition is within reach.

\subsection{Policy implications} 
A benchmark orbital treatment fee with a scheduled step-down, similar to China's 0.65 CNY/kWh feed-in tariff, can create a revenue floor without a permanent subsidy \cite{shen2024}. The weighbridge analogy supplies the second, as the commercial SSA catalogs must be licensed and standardized as settlement infrastructure for charging. Without measurement, there is no commodity. Post-mission disposal (PMD) bonds extend the MSW treatment-fee precedence into space, converting an obligation into a valuable asset. And the WtE record carries a warning that planned capacity ahead of verified demand produces overbuild \cite{shapiro2020}, so orbital policy should allocate planned removal capacity to licensed, insured, and fee-paying demand rather than to declarations. The \texttt{S3+1} non-profit is the natural breeding ground for these standards 
consistent with its role as the trust infrastructure of the HKSAR-based framework. The space orbit lacks a public collection network, States do not yet pay per tonne of debris, and the tariff levels cannot be transferred. What transfers is the logic that a coupled incentive stack drives unit cost down until a liability becomes an asset — and that, on the terrestrial record, is how ``making space sustainability pay'' actually happens.

\section{Conclusion} \label{sec8}
The technology for debris observation and removal has reached the commercial demonstration stage. The economics have not. The orbital commons will be governed one way or another. The question is whether the governing structure couples deliberately - binding disposal mandates, forensic SSA, insurance-linked and pre-funded instruments, and liability rules that attribute and allocate cost or lets the market drift. As a result of such integration, space sustainability will become a service category rather than a subsidy, and remediation ``makes pay'' in the only sense that contributes to the banking, insurance, and investment business. Otherwise, in the long term, the commons will keep getting priced by collision statistics, not by stewardship, and the 2030s will bear the cost in lost capability, lost revenue, and lost environment.

\section*{Author Contributions}
J.~C., P.~S.~P., S.~S., and A.~R. contributed to the manuscript writing. M.~S. contributed to the planning and supervision.








\bibliography{reference}  

@misc{esa2026,
title = "{Space Debris by the Numbers (MASTER 8) --- ESA(2026)}",
author = {},
howpublished = {Available online: \url{https://sdup.esoc.esa.int/discosweb/statistics/embed/bythenumbers}},
note         = {accessed on 30 October 2026},
}

@misc{ep2026,
title = {Orbital congestion: A growing threat to space activities. --- {European Parliament (2026)}}, 
howpublished = {Available online: \url{https://www.europarl.europa.eu/RegData/etudes/BRIE/2026/795713/ECTI_BRI(2026)795713_EN.pdf}},
note         = {accessed on 31 October 2026},
}

@misc{esa2025,
title = {ESA's Annual Space Environment Report (Issue 9.1) --- {ESA (2025)}}, 
howpublished = {Available online: \url{https://www.sdo.esoc.esa.int/publications/Space_Environment_Report_I9R1_20251021.pdf}},
note = {accessed on 31 October 2026},
}

@misc{kt2026,
title = {The 300,000 Swerves Keeping Low Earth Orbit From Crashing --- {KeepTrack (2026)}},
howpublished = {Available online: \url{https://www.keeptrack.space/deep-dive/collision-avoidance-conjunction-screening}},
note = {accessed on 31 October 2026},
}

@ARTICLE{1978JGR....83.2637K,
       author = {{Kessler}, Donald J. and {Cour-Palais}, Burton G.},
        title = "{Collision frequency of artificial satellites: The creation of a debris belt}",
      journal = {Journal of Geophysical Research},
         year = 1978,
        month = jun,
       volume = {83},
       number = {A6},
        pages = {2637-2646},
          doi = {10.1029/JA083iA06p02637},
          note = "\url{http://dx.doi.org/10.1029/JA083iA06p02637}",
       adsurl = {https://ui.adsabs.harvard.edu/abs/1978JGR....83.2637K}
}

@ARTICLE{1968Sci...162.1243H,
       author = {{Hardin}, Garrett},
        title = "{The Tragedy of the Commons}",
      journal = {Science},
         year = 1968,
        month = dec,
       volume = {162},
       number = {3859},
        pages = {1243-1248},
          doi = {10.1126/science.162.3859.1243},
          note = "\url{http://dx.doi.org/10.1126/science.162.3859.1243}",
       adsurl = {https://ui.adsabs.harvard.edu/abs/1968Sci...162.1243H}
}

@ARTICLE{1998Sci...280..682H,
       author = {{Hardin}, Garrett},
        title = "{ESSAYS ON SCIENCE AND SOCIETY: Extensions of ``The Tragedy of the Commons''}",
      journal = {Science},
         year = 1998,
        month = may,
       volume = {280},
        pages = {682},
        doi = {10.1126/science.280.5364.682},
        note = "\url{http://dx.doi.org/10.1126/science.280.5364.682}",
       adsurl = {https://ui.adsabs.harvard.edu/abs/1998Sci...280..682H}
}

@article{weeden2012,
title = {Taking a common-pool resources approach to space sustainability: A framework and potential policies},
journal = {Space Policy},
volume = {28},
number = {3},
pages = {166-172},
year = {2012},
issn = {0265-9646},
doi = {https://doi.org/10.1016/j.spacepol.2012.06.004},
note = "\url{https://www.sciencedirect.com/science/article/pii/S0265964612000604}",
author = {{Weeden}, Brian C. and {Chow}, Tiffany}
}

@misc{nse2026a,
title = {Commercial Ground-Based Space Warfare Systems Reshaping the Orbital Contest --- {New Space Economy (2026a)}}, 
howpublished = {Available online: \url{https://newspaceeconomy.ca/2026/04/17/commercial-ground-based-space-warfare-systems-reshaping-the-orbital-contest/}},
note = {accessed on 31 October 2026},
}

@misc{nse2026b,
title = {What Does NorthStar's Public-Market Debut Reveal About Investment in Space-Domain Awareness? --- {New Space Economy (2026b)}},  howpublished = {Available online: \url{https://newspaceeconomy.ca/2026/10/04/what-does-northstars-public-market-debut-reveal-about-investment-in-space-domain-awareness/}},
note = {accessed on 31 October 2026},
}

@misc{leolab2025,
title = {LeoLabs Receives Contract from the U.S. Department of Commerce and U.S. Space Force To Jointly License Its Object Catalog for Space Safety and Security Missions},
howpublished = {Available online: \url{https://leolabs.space/press/osc-jco-license-leolabs-object-catalog/}},
note = {Accessed on 31 October 2026},
}

@misc{iaf2026,
title = {Paper of the IAF Space Traffic Management Working Group --- {IAF (2026)}},
howpublished = {Available online: \url{https://www.unoosa.org/documents/pdf/copuos/lsc/2026/STM/IAF_STM.pdf}},
note = {Accessed on 31 October 2026},
}

@misc{esa2024,
title = {ClearSpace-1 in-orbit demonstration mission [Clean Space Days 2024, ESA/ESTEC]}, 
howpublished = {Available online: \url{https://indico.esa.int/event/516/contributions/9983/attachments/6342/10932/ClearSpace-1%20IOD%20Mission_Clean%20Space%20Days.pdf},
note = {Accessed on 31 October 2026},
}}

@misc{defen2026,
title = {Northrop Grumman launches in-space servicing satellites for life-extension missions.},
howpublished = {Available online: \url{https://defensescoop.com/2026/07/22/northrop-grumman-launches-mission-robotic-vehicle-jetpack-pods/}},
note = {Accessed on 31 October 2026},
}

@INPROCEEDINGS{2013ESASP.723E..85W,
       author = {{Wormnes}, K. and {Le Letty}, R. and {Summerer}, L. and {Schonenborg}, R. and {Dubois-Matra}, O. and {Luraschi}, E. and {Cropp}, A. and {Krag}, H. and {Delaval}, J.},
        title = "{ESA Technologies for Space Debris Remediation}",
    booktitle = {6th European Conference on Space Debris},
         year = 2013,
       editor = {{Ouwehand}, L.},
       series = {ESA Special Publication},
       volume = {723},
        month = aug,
          eid = {85},
        pages = {85},
        note = "\url{https://conference.sdo.esoc.esa.int/proceedings/sdc6/paper/116/SDC6-paper116.pdf}",
       adsurl = {https://ui.adsabs.harvard.edu/abs/2013ESASP.723E..85W}
}

@misc{sn2021,
title = {Privately funded mission takes off to begin space debris cleanup trials ---{Spaceflight Now. (2021, March 22).} },
howpublished = {Available online: \url{https://spaceflightnow.com/2021/03/22/privately-funded-mission-takes-off-to-begin-space-debris-cleanup-trials/}},
note = {Accessed on 31 October 2026},
}

@ARTICLE{2026AdSpR..7712265A,
       author = {{Alnaqbi}, F. and {Biktimirov}, S. and {Gaias}, G.},
        title = "{Conceptual design and analysis of a NanoTug swarm for active debris removal}",
      journal = {Advances in Space Research},
         year = 2026,
        month = jun,
       volume = {77},
       number = {12},
        pages = {12265-12285},
          doi = {10.1016/j.asr.2026.04.030},
          note = "\url{http://dx.doi.org/10.1016/j.asr.2026.04.030}",
archivePrefix = {arXiv},
       eprint = {2604.19214},
 primaryClass = {astro-ph.EP},
       adsurl = {https://ui.adsabs.harvard.edu/abs/2026AdSpR..7712265A}
}

@article{abott2025,
author = {{Abbott}, Jake J.},
title = {Roboticists are grappling with space debris},
journal = {Science Robotics},
volume = {10},
number = {103},
pages = {eadt5685},
year = {2025},
doi = {10.1126/scirobotics.adt5685},
note = "\url{http://dx.doi.org/10.1126/scirobotics.adt5685}",
URL = {https://www.science.org/doi/abs/10.1126/scirobotics.adt5685},
eprint = {https://www.science.org/doi/pdf/10.1126/scirobotics.adt5685}}

@article{WILLIAMSROGERS20255265,
title = {Optimal placement and coordinated scheduling of distributed space-based lasers for orbital debris remediation},
journal = {Advances in Space Research},
volume = {76},
number = {9},
pages = {5265-5293},
year = {2025},
issn = {0273-1177},
note = {\url{https://doi.org/10.1016/j.asr.2025.07.093}},
url = {https://www.sciencedirect.com/science/article/pii/S027311772500849X},
author = {David O. {Williams Rogers} and Matthew C. Fox and Paul R. Stysley and Hang Woon Lee}
}

@misc{scio2022,
title = {China develops de-orbiting sail to manage space debris -- {SCIO (2022)}}, 
howpublished = {Available online: \url{http://english.scio.gov.cn/m/chinavoices/2022-11/21/content_78529357.htm}},
note = {Accessed on 3 November 2026},
}

@inproceedings{sdc9paper121,
  author    = {Toru {Yamamoto} and  Hiroyuki Okamoto and Makoto Tomitaka and Ryo Nakamura and Yu Takeuchi},
  title     = {Advancing Active Debris Removal: Achievements and Prospects in the Commercial Removal of Debris Demonstration Project},
  booktitle = {Proceedings of the 9th {ESA} Space Debris 
               Conference ({SDC9})},
  address   = {Darmstadt, Germany},
  month     = {April},
  year      = {2025},
  editor    = {{Lemmens} S. and {Flohrer} T. and {Schmitz} F.},
  publisher = {European Space Agency ({ESA})},
  number    = {121},
  pages     = {1--11},
  note      = {Available online: \url{https://conference.sdo.esoc.esa.int/proceedings/sdc9/paper/121/SDC9-paper121.pdf} (accessed on 3 November 2026)},
}

@misc{sn2022,
title = {China's Shijian-21 towed dead satellite to a high graveyard orbit -- {SpaceNews (2022)}}, 
howpublished = {Available online: \url{https://spacenews.com/chinas-shijian-21-spacecraft-docked-with-and-towed-a-dead-satellite/}},
note = {Accessed on 3 November 2026},
}

@misc{esa2023,
title = {Zero Debris Charter: Towards a safe and sustainable space environment -- {ESA/ESOC 2023}},
howpublished = {Available online: \url{https://esoc.esa.int/sites/default/files/Zero_Debris_Charter_EN.pdf}},
note = {Accessed on 3 November 2026},
}

@misc{fcc2022,
title = {Second Report and Order, Space Innovation; Mitigation of Orbital Debris in the New Space Age (FCC 22-74; 47 CFR 25.283(e)) -- {FCC (2022)}},
howpublished = {Available online: \url{https://www.federalregister.gov/documents/2024/08/09/2024-17093/space-innovation-mitigation-of-orbital-debris-in-the-new-space-age}},
note = {Accessed on 3 November 2026},
}

@misc{orbin2026,
title = {LeoLabs Wins \$20.68M Space Force Scout Contract -- {Orbital Intel (2026)}},
howpublished = {Available online: \url{https://orbital-intel.com/news/leolabs-20-68m-space-force-scout-contract/}},
note = {Accessed on 3 November 2026},
}

@misc{lsa2026,
title = {ClearSpace's Phoenix mission to extend satellite life directly in orbit gets the green light  --- {Luxembourg Space Agency (2026)}},
howpublished = {Available online: \url{https://space-agency.public.lu/en/news-events/news/2026/clearspaces-phoenix-mission-to-extend-satellite-life-directly-in-orbit-gets-the-green-light.html}},
note = {Accessed on 3 November 2026},
}

@misc{smr2026,
title = {In-orbit services market to generate US\$3 bil. in revenue over the next decade -- {Satellite Markets \& Research (2026-08-10)}},
howpublished = {Available online: \url{https://www.satellitemarkets.com/market-trends/orbit-services-market-generate-us3-bil-revenue-over-next-decade}},
note = {Accessed on 3 November 2026},
}

@misc{wtw2025,
title = {Insurance Marketplace Realities 2026; Aviation \& Space -- {WTW (2025-10-02)}},
howpublished = {Available online: \url{https://www.wtwco.com/en-us/insights/2025/10/insurance-marketplace-realities-2026-aviation-and-space}},
note = {Accessed on 3 November 2026},
}

@misc{inasia2026,
title = {China's space insurance consortium writes US\$1.5bn in cover in first year -- {(Re)inAsia (2026-05-11)}}, 
author = {Jessica {Li}},
howpublished = {Available online: \url{https://reinasia.com/chinas-space-insurance-consortium-writes-us1-5bn-in-cover-in-first-year/}},
note = {Accessed on 3 November 2026},
}

@misc{bd2026,
title = {SpainSat Loss Delivers Shock to Space Insurance Market --- {Insurance News BD (2026-05-16)}},
howpublished = {Available online: \url{https://eng.insurancenewsbd.com/article/news/6578}},
note = {Accessed on 3 November 2026},
}

@misc{wef2024,
title = {Financial Space Debris Mechanisms Workshop – Summary Notes --{WEF (2024-06-06)}},
author = {Nikolai {Khlystov}},
howpublished = {Available online: \url{https://www.mainelli.org/wp-content/uploads/2024/07/Summary-Notes_-Financial_Space_Debris.pdf}},
note = {Accessed on 3 November 2026},
}

@misc{un1972,
title = {Convention on International Liability for Damage Caused by Space Objects. United Nations Treaty Series, vol. 961. 1972}, 
howpublished = {Available online: \url{https://treaties.un.org/untc/Pages/doc/Publication/UNTS/Volume%20961/volume-961-I-13810-English.pdf}},
note = {Accessed on 3 November 2026},
}

@misc{unoosa1972,
title = {Convention on International Liability for Damage
Caused by Space Objects (A/RES/2777(XXVI)) -- {UNOOSA (1972)}},
howpublished = {Available online: \url{https://www.unoosa.org/oosa/en/ourwork/spacelaw/treaties/liability-convention.html}},
note = {Accessed on 3 November 2026},
}

@misc{slj2026,
title = {The FCC Five Year Deorbit Rule: Who It Binds and What It Requires --- {Space Law Journal (2026)}}, 
howpublished = {Available online: \url{https://spacelawjournal.com/fcc-five-year-deorbit-rule/}},
note = {Accessed on 3 November 2026},
}

@misc{ukgov2026,
title = {Space regulatory financial tools package: Including government response to the liability, insurance, charging and space sustainability consultation --- {GOV.UK (2026-09-08)}},
howpublished = {Available online: \url{https://www.gov.uk/government/consultations/consultation-on-orbital-liabilities-insurance-charging-and-space-sustainability/outcome/space-regulatory-financial-tools-package-including-government-response-to-the-liability-insurance-charging-and-space-sustainability-consultation}},
note = {Accessed on 3 November 2026},
}

@misc{legifr2008,
title = {Law No. 2008-518 of 3 June 2008 relating to space operations --- {Légifrance (2008)}},
howpublished = {Available online: \url{https://www.legifrance.gouv.fr/jorf/id/JORFSCTA000018931386}},
note = {Accessed on 3 November 2026},
}

@misc{unoosa2019,
title = {Guidelines for the Long-Term Sustainability of Outer Space Activities (A/74/20, Annex II) -- {UNOOSA 2019}}, 
howpublished = {Available online: \url{https://www.unoosa.org/unoosa/oosadoc/data/documents/2019/a/a7420_0.html}},
note = {Accessed on 3 November 2026},
}

@misc{unoosa2026,
title = {Compendium of space debris mitigation standards adopted by States and international organizations (IADC Guidelines 2007, updated 2025) --- {UNOOSA (2026)}}, 
howpublished = {Available online: \url{https://www.unoosa.org/oosa/en/ourwork/topics/space-debris/compendium.html}},
note = {Accessed on 3 November 2026},
}

@misc{iso2023,
title = {ISO 24113:2023 – Space systems – Space debris mitigation requirements (4th ed.)},
howpublished = {Available online: \url{https://www.iso.org/es/contents/data/standard/08/34/83494.html}},
note = {Accessed on 3 November 2026},
}

@misc{itu2023,
title = {Resolution 8 (WRC-23); Resolution ITU-R 74 (RA-23) -- {ITU (2023)}},
howpublished = {Available online: \url{https://www.itu.int/en/ITU-R/space/Documents/RES8%20%28WRC-23%29.pdf}},
note = {Accessed on 3 November 2026},
}

@misc{itu2024,
title = {Circular Letter CA/272: Deorbit and/or disposal strategies for non-GSO space stations using frequency assignments currently recorded in the ITU MIFR, or under coordination or notification processes --- {ITU (2024)}},
howpublished = {Available online: \url{https://www.itu.int/space-sustainability/wp-content/uploads/sites/40/2024/03/CA-272-E.pdf}},
note = {Accessed on 3 November 2026},
}

@misc{ssr2025,
title = {Space Sustainability Rating – Promoting Sustainable Behavior of Space Actors -- {SSR (2025)}},
howpublished = {Available online: \url{https://spacesustainabilityrating.org}},
note = {Accessed on 3 November 2026},
}

@ARTICLE{shen2024,
AUTHOR={Shen, Ye  and Xu, Min  and Cui, Caiyun  and Xia, Bo  and Skitmore, Martin  and Moorhead, Matthew  and Liu, Yong },
TITLE={Impact of environmental regulation on government subsidies of public-private partnership waste-to-energy incineration projects: evidence from 66 cities in China},
JOURNAL={Frontiers in Environmental Science},
VOLUME={Volume 11 - 2023},
YEAR={2024},
URL={https://www.frontiersin.org/journals/environmental-science/articles/10.3389/fenvs.2023.1288851},
note = {\url{https://www.frontiersin.org/journals/environmental-science/articles/10.3389/fenvs.2023.1288851}},
DOI={10.3389/fenvs.2023.1288851},
ISSN={2296-665X}}

@misc{iea2022,
title = {2021 biomass power generation project construction work program [Policies Database, No. 14306] --- {International Energy Agency (2022)}},
howpublished = {Available online: \url{https://www.iea.org/policies/14306-national-development-and-reform-commission-ministry-of-finance-national-energy-administration-2021-biomass-power-generation-project-construction-work-program}},
note = {Accessed on 3 November 2026},
}

@misc{wb2010,
title = {Municipal solid waste treatment project, Wenzhou, China (Public–Private Partnership case study) --- {World Bank (2010)}},
howpublished = {Available online: \url{https://ppp.worldbank.org/municipal-solid-waste/municipal-solid-waste-treatment-project-wenzhou-china}},
note = {Accessed on 3 November 2026},
}

@article{cui2026,
  author    = {Cui, Jicui and Cui, Ying and Li, Jiyang and Gao, Xiaofeng and 
               Wei, Wendong and Chen, Yuan and Ma, Wenchao and Zhu, Nanwen and 
               Geng, Yong and Zhao, Youcai and Lou, Ziyang},
  title     = {Efficiency hierarchy and optimization of waste incineration in 
               {China} to balance disposal and energy supply},
  journal   = {Nature Communications},
  year      = {2026},
  volume    = {17},
  number    = {1},
  pages     = {3069},
  doi       = {10.1038/s41467-026-69897-w},
  url       = {https://doi.org/10.1038/s41467-026-69897-w},
  note      = {\url{https://doi.org/10.1038/s41467-026-69897-w}},
  issn      = {2041-1723},
  date      = {2026-02-23},
}

@article{shapiro2020,
title={Municipal solid waste available to the Chinese energy sector - Provincial projections to 2050},
  author={Sara, Shapiro-Bengtsen and Frits M{\o}ller, Andersen and Marie, M{\"u}nster and Le-Le, Zou},
  journal={Waste management},
  year={2020},
  volume={112},
  pages={52-65},
  url={https://api.semanticscholar.org/CorpusID:219550818},
  doi={10.1016/j.wasman.2020.05.014},
  note={\url{http://dx.doi.org/10.1016/j.wasman.2020.05.014}}
  }

@article{adilov2015,
  title={An Economic Analysis of Earth Orbit Pollution},
  author={Nodir, Adilov and Peter J., Alexander and Brendan M., Cunningham},
  journal={Environmental and Resource Economics},
  year={2015},
  volume={60},
  pages={81-98},
  DOI={10.1007/S10640-013-9758-4},
  url={https://api.semanticscholar.org/CorpusID:153395142},
  note={\url{https://api.semanticscholar.org/CorpusID:153395142}}
}

@ARTICLE{bongers2023,
       author = {{Bongers}, Anel{\'\i} and {Torres}, Jos{\'e} L.},
        title = "{Orbital debris and the market for satellites}",
      journal = {Ecological Economics},
         year = 2023,
        month = jul,
       volume = {209},
          eid = {107831},
        pages = {107831},
          doi = {10.1016/j.ecolecon.2023.107831},
          note = "\url{http://dx.doi.org/10.1016/j.ecolecon.2023.107831}",
       adsurl = {https://ui.adsabs.harvard.edu/abs/2023EcoEc.20907831B}
}

@ARTICLE{rao2020,
       author = {{Rao}, Akhil and {Burgess}, Matthew G. and {Kaffine}, Daniel},
        title = "{Orbital-use fees could more than quadruple the value of the space industry}",
      journal = {Proceedings of the National Academy of Science},
         year = 2020,
        month = jun,
       volume = {117},
       number = {23},
        pages = {12756-12762},
          doi = {10.1073/pnas.1921260117},
          note= {\url{http://dx.doi.org/10.1073/pnas.1921260117}},
       adsurl = {https://ui.adsabs.harvard.edu/abs/2020PNAS..11712756R}
}

@misc{nasa2024,
title = {Cost and benefit analysis of mitigating, tracking, and remediating orbital debris --- {NASA (2024)}},
howpublished = {Available online: \url{https://ntrs.nasa.gov/citations/20240003484}},
note = {Accessed on 3 November 2026},
}

@TechReport{bongers2025,
type={Space Economics Working Papers},
institution={Institute for Space Economics},
author={Aneli Bongers and Jose L. Torres},
title={Optimal Active Debris Removal Policy in the Long-run},
year={2025},
month={Jul},
number={07-2025},
doi={None},
url={https://ideas.repec.org/p/bhw/wpaper/07-2025.html},
note={\url{https://spaceeconomics.org/wp-content/uploads/2025/07/SEWP-07-2025.pdf}}
}

@misc{ssm2011,
title = {DARPA's PHOENIX: Scavenging Defunct GEO Satellites --- {Space Safety Magazine (2011)}},
howpublished = {Available online: \url{http://www.spacesafetymagazine.com/press-release/darpas-phoenix-scavenging-defunct-geo-satellites/}},
note = {Accessed on 3 November 2026},
}

@misc{iaf2022,
title = {Solving space debris by using it as a space resource for the emerging industrial space economy (IAC-22, A6, IPB, 2, x73756) --- {IAF (2022)}},
howpublished = {Available online: \url{https://iafastro.directory/iac/paper/id/73756/abstract-pdf/IAC-22,A6,IPB,2,x73756.brief.pdf}},
note = {Accessed on 3 November 2026},
}

@misc{iaf2025,
title = {XGS space waste recycling station: a sustainable resource manufacturing in orbit (IAC-25, C2, IP, 90, x94786) -- {IAF (2025)}},
howpublished = {Available online: \url{https://iafastro.directory/iac/paper/id/94786/abstract-pdf/IAC-25,C2,IP,90,x94786.brief.pdf}},
note = {Accessed on 3 November 2026},
}

@article{leonard2023,
title = {Viability of a circular economy for space debris},
journal = {Waste Management},
volume = {155},
pages = {19-28},
year = {2023},
issn = {0956-053X},
doi = {https://doi.org/10.1016/j.wasman.2022.10.024},
url = {https://www.sciencedirect.com/science/article/pii/S0956053X22005104},
note = {\url{https://www.sciencedirect.com/science/article/pii/S0956053X22005104}},
author = {Ryan, Leonard and Ian D., Williams}
}

@ARTICLE{lenkart2026,
AUTHOR={Lenkart, Madeline and Biller, Stephan},
TITLE={Space debris material sourcing for in-space manufacturing: a quantitative evaluation framework},
JOURNAL={Frontiers in Space Technologies},
VOLUME={Volume 6 - 2025},
YEAR={2026},
URL={https://www.frontiersin.org/journals/space-technologies/articles/10.3389/frspt.2025.1690460},
note={\url{https://www.frontiersin.org/journals/space-technologies/articles/10.3389/frspt.2025.1690460}},
DOI={10.3389/frspt.2025.1690460},
ISSN={2673-5075}}

@article{morin2021,
author = {Morin, Jean-Frédéric and Richard, Benjamin},
title = {Astro-Environmentalism: Towards a Polycentric Governance of Space Debris},
journal = {Global Policy},
volume = {12},
number = {4},
pages = {568-573},
doi = {https://doi.org/10.1111/1758-5899.12950},
url = {https://onlinelibrary.wiley.com/doi/abs/10.1111/1758-5899.12950},
note={\url{https://onlinelibrary.wiley.com/doi/abs/10.1111/1758-5899.12950}},
eprint = {https://onlinelibrary.wiley.com/doi/pdf/10.1111/1758-5899.12950},
year = {2021}
}

@article{asaad2026,
author = {Abdul-Hamid, Asaad and Pearl, Brycen D. and Lee, Hang Woon and Chen, Hao},
title = {Space Logistics Analysis and Incentive Design for Commercialization of Orbital Debris Remediation},
journal = {Journal of Spacecraft and Rockets},
volume = {63},
number = {1},
pages = {218-230},
year = {2026},
doi = {10.2514/1.A36465},
URL = {https://doi.org/10.2514/1.A36465},
eprint = {https://doi.org/10.2514/1.A36465}
}

@misc{cd202609,
title = {HK’s five-year plan can propel NewSpace economy -- {China Daily (2026-09-18)}},
author = {Quentin {Parker}},
howpublished = {Available online: \url{https://www.chinadailyhk.com/upload/main/pdf/2026/09/18/204ff6ba9308e2a998ce6b9ac9cfe71e.pdf}},
note = {Accessed on 7 November 2026},
}

@misc{hk5,
title = {The First Five-Year Plan for Economic and Social Development of the Hong Kong Special Administrative Region (2026-2030) -- {HKSAR GOV (2026-09)}},
howpublished = {Available online: \url{https://www.hk5yplan.gov.hk/2026-2030/en/index.html}},
note = {Accessed on 7 November 2026},
}

\begin{landscape}
\begin{table}[htp!]
\caption{Reference ADR and on-orbit servicing missions and their 2026 status.}
\label{op_stat}
\begin{center}
\scriptsize
\begin{tabular}{|c|c|c|c|c|}
\hline
\textbf{Mission} & \textbf{Lead} & \textbf{Capture/Service} & \textbf{Target} & \textbf{Status} \\
\textbf{} & \textbf{(Agency/Company)} & \textbf{technology} & \textbf{} & \textbf{(2026)} \\\hline
RemoveDEBRIS &University of Surrey/EU & Net+harpoon & CubeSat mock debris & 2018; demonstrated net \& harpoon capture in LEO \\\hline
ELSA-d & Astroscale & Magnetic docking plate & Servicer's own client& 2021; first commercial magnetic capture demo \\\hline
MEV-1/MEV-2 & SpaceLogistics & Mating adaptor, & Intelsat 901; & GEO life-extension leased service since 2020--21;\\
 & (Northrop Grumman) & life extension & Intelsat 10-02 & MEV-1 separated from Intelsat 901 in 2025 \\\hline
SJ-21 & China Aerospace & Net capture, propulsive tow & Defunct Beidou-2 G2 & 2022; towed target to super-graveyard orbit in GEO \\\hline
ADRAS-J & Astroscale / JAXA & RPO inspection & H-IIA upper stage & 2024--2025; first close inspection of large debris;\\
 & CRD2 Phase I & (no capture) & (${\sim}$3~t) & deorbit initiated \\\hline
ADRAS-J2 & Astroscale / JAXA & Capture \& removal & Same H-IIA upper stage & Under development \\
& CRD2 Phase II & & & \\\hline
ClearSpace-1 & ESA / ClearSpace & Four-arm rigid capture & Defunct PROBA-1 & Revised 2024; launch planned 2028; \\
& &  & & plus Phoenix GEO life-extension contract (2026) \\\hline
MRV+MEPs & SpaceLogistics/  & Robotic arms; & GEO satellites & Launched July 2026;\\
 &  & MEP jetpack pods & & inspection/repair/pod installation\\\hline
\end{tabular}\\
Sources: DefenseScoop (2026)\cite{defen2026}; ESA (2024)\cite{esa2024}; JAXA (2025)\cite{sdc9paper121}; SCIO (2022)\cite{scio2022}; Spaceflight Now (2021)\cite{sn2021}; SpaceNews (2022)\cite{sn2022}.
\end{center}
\end{table}

\begin{table}[htp!]
\caption{Commercial sustainability business models (2015–2026).}
\label{tab_biz}
\begin{center}
\scriptsize
\begin{tabular}{|c|c|c|c|c|}
\hline
\textbf{Model} & \textbf{Representative players} & \textbf{Capture/Service} & \textbf{Revenue logic} & \textbf{Maturity} \\\hline
SSA DaaS  (subscription)& 
LeoLabs, ExoAnalytic, Slingshot, Karman+ & 
Operators, insurers, agencies &
Recurring data subscriptions + tasking &
Commercial \\\hline
SDA/government contracts &
LeoLabs, NorthStar, True Anomaly &
USSF, OSC, allied governments & 
Multi-year anchor contracts & 
Scaling \\\hline 
ADR mission procurement & 
Astroscale, ClearSpace &
JAXA, ESA, UKSA, USSF & 
Fixed-price institutional missions &
First-of-kind \\\hline 
Disposal-as-a-service / EOL & 
D-Orbit, Starfish, Exotrail-Astroscale & 
LEO operators, launchers & 
Per-satellite disposal or bundled launch service & 
Emerging \\\hline 
Life-extension leasing & 
SpaceLogistics (MEV/MEP) & 
GEO operators & 
Lease of mission extension time & 
Commercial \\\hline 
Refueling infrastructure &
Orbit Fab &
Servicers, operators &
Fuel purchase / subscription &
Pre-commercial \\\hline 
Insurance-linked  &
Insurers + servicers &
Operators, ADR providers & 
Premiums linked to verified disposal &
Pilot/concept \\
sustainability products &
(concept phase) &
 & 
&
\\\hline 
\end{tabular}
\end{center}
\end{table}

\end{landscape}


\end{document}